\documentclass[journal]{IEEEtran}
\usepackage{graphicx}
\usepackage{subfigure}
\usepackage{subeqnarray}
\usepackage{bm}
\usepackage{amsfonts}
\usepackage{amsthm}
\usepackage{amssymb}
\usepackage{makecell}
\usepackage{calc}
\usepackage{color}
\usepackage{mathrsfs}
\usepackage{flushend}
\usepackage{url}
\usepackage{caption}
\usepackage{subfloat}
\usepackage{color}
\usepackage{textcomp}
\usepackage{stfloats}

\usepackage[outdir=./]{epstopdf}

\usepackage{cite}

\usepackage{booktabs}

\newtheorem{remark}{Remark}

\ifCLASSINFOpdf
\else
\fi
\usepackage{amsmath}
\usepackage{amsthm}

\usepackage{algorithm}
\usepackage{algorithmic}
\ifCLASSOPTIONcompsoc
 \usepackage[caption=false,font=normalsize,labelfont=sf,textfont=sf]{subfig}
\else
 \usepackage[caption=false,font=footnotesize]{subfig}
\fi
\begin{document}
\captionsetup[figure]{labelformat={default},labelsep=period,name={Fig.}}
\title{Sensing-Assisted Eavesdropper Estimation: An ISAC Breakthrough in Physical Layer Security}
%
%
%

\author{Nanchi Su,~\IEEEmembership{Graduate~Student~Member,~IEEE}, Fan Liu,~\IEEEmembership{Member,~IEEE}, Christos Masouros,~\IEEEmembership{Senior~Member,~IEEE}

\thanks{{This work was supported in part by the Engineering and Physical Sciences Research Council (EPSRC) under Grant EP/S028455/1, in part by the National Natural Science Foundation of China under Grant 62101234 and Grant U20B2039, in part by the Young Elite Scientist Sponsorship Program by CAST under Grant No. YESS20210055, and in part by the China Scholarship Council (CSC). \textit{(Corresponding author: Fan Liu.)}}}
\thanks{N. Su and C. Masouros are with the Department of Electronic and Electrical Engineering, University College London, London WC1E 7JE, U.K. (e-mail: nanchi.su.18@ucl.ac.uk, chris.masouros@ieee.org).}
\thanks{F. Liu is with the Department of Electrical and Electronic Engineering, Southern University of Science and Technology, Shenzhen 518055, China (e-mail: liuf6@sustech.edu.cn).}
}

\maketitle

\begin{abstract}
   In this paper, we investigate the sensing-aided physical layer security (PLS) towards Integrated Sensing and Communication (ISAC) systems. A well-known limitation of PLS is the need to have information about potential eavesdroppers (Eves). The sensing functionality of ISAC offers an enabling role here, by estimating the directions of potential Eves to inform PLS. In our approach, the ISAC base station (BS) firstly emits an omni-directional waveform to search for potential Eves' directions by employing the combined Capon and approximate maximum likelihood (CAML) technique. Using the resulting information about potential Eves, we formulate secrecy rate expressions, that are a function of the Eves' estimation accuracy. We then formulate a weighted optimization problem to simultaneously maximize the secrecy rate and minimize the CRB with the aid of the artificial noise (AN), and minimize the CRB of targets'/Eves' estimation. By taking the possible estimation errors into account, we enforce a beampattern constraint with a wide main beam covering all possible directions of Eves. This implicates that security needs to be enforced in all these directions. By improving estimation accuracy, the sensing and security functionalities provide mutual benefits, resulting in improvement of the mutual performances with every iteration of the optimization, until convergence. Our results avail of these mutual benefits and reveal the usefulness of sensing as an enabler for practical PLS.
\end{abstract}

\begin{IEEEkeywords}
  Integrated Sensing and Communication system, sensing aided physical layer security, Cram\'{e}r-Rao Bound, secrecy rate, artificial noise.
\end{IEEEkeywords}

%
\IEEEpeerreviewmaketitle

\section{Introduction}
%
%
%
%
\subsection{Background and Motivation}
\IEEEPARstart{A}{s} the 5G wireless networks are being rolled-out worldwide, emerging applications, such as connected cars, smart factories, digital twins,  highlight the limitations of existing network infrastructures \cite{you2021towards}. These applications demand both increasingly high quality of commmunication as well as high accuracy and robustness of sensing, it is well-recognized that the cooperation and co-design between communication and radar systems will play a significant role in the upcoming beyond 5G (B5G) and 6G eras.
\\\indent At the early stage of the radar-communication (RadCom) system studies, the two systems were conceived to spectrally coexist with each other, thus to ease the severe competition over the scarce spectrum resources \cite{feng2020joint, 8828016}. In the forthcoming B5G/6G eras, radio sensing and communications (S\&C) are both evolving towards higher frequency bands and large-scale antenna arrays, which leads to striking similarities between S\&C systems in terms of hardware architecture, channel charateristics, and information processing pipeline \cite{liu2022integrated}. In light of this, the research on the coexistence of radar and communication systems has involved into dual-functional radar communication (DFRC) systems. The joint design of the S\&C operations, in the form of Integrated Sensing and Communications (ISAC), have been initially proposed in \cite{cui2021integrating}. ISAC systems are expected to achieve higher spectral and energy efficiencies, but most importantly, promote a new paradigm of integration for attaining mutual benefits from a co-design perspective, wherein the S\&C functionalities can mutually assist each other. Benefiting from these two advantages, applications of ISAC have been extended to numerous emerging areas, including smart manufacturing, environmental monitoring, vehicular networks, as well as indoor services such as human activity recognition.
\\\indent With the evolution of cellular networks, the security in mmWave ISAC systems is facing with great challenges because of the shared use of the spectrum and the broadcasting nature of wireless transmission \cite{wei2022toward}. On one hand, the Rician channels are widely employed in mmWave frequencies, containing the line of sight (LoS) component, which results in the inescapable correlation with the sensing channel. This is different from conventional physical layer security (PLS) studies in communication systems with the independent and identically distributed assumption between legitimate user channels and intercept channels \cite{liu2017enhancing,qin2016physical,hamamreh2018classifications}. On the other hand, in the dual-functional waveform design, the confidential information intended for communication users (CUs) is embedded in radar probing signals, which is susceptible of being eavesdropped by the target of interest. In this case, a unique and interesting conflict arises from the radar functionality side. To be specific, the power is expected to be focused towards targets of interest to improve the detectability, while the useful signal information has to be protected from being intercepted by the targets, which are acknowledged as Eves, as each of them is reckoned as a potential eavesdropper (Eve).
\\\indent To secure the confidential information in ISAC systems, existing approaches can be generally divided into following categories, i.e., 1) Cryptography and 2) PLS. Conventionally, the security of communication systems is regarded as an independent feature and addressed at upper layers of protocol stack by deploying cryptographic technologies. The studies of cryptography commonly assume that the physical layer provides an error-free link \cite{melki2019survey}, while the wireless links are vulnerable to attacks in practice, which would result in a high risk of information leakage. It is worth pointing out that 5G has already been a large-scale heterogeneous network with multiple levels and weakly-structured architectures, which makes it difficult to distribute and manage secret keys \cite{sun2017physical}. Also, complicated encryption/decryption algorithms cannot be straightforwardly applied considering the power consumption in 5G networks. Furthermore, even if the data is encrypted, the detection of a wireless link from a potential eavesdropper can reveal critical information. In contrast to complex cryptographic approaches, signal processing operations of PLS are usually simple with little additional overheads. A major limitation of PLS is the need to obtain some information for the potential Eves. This ranges from full CSI, to an SNR estimate of the Eve's link, or the Eve's direction as a minimum. This difficult-to-obtain information often renders PLS impractical.
\\\indent More relevant to this work, security in ISAC systems was initially studied in \cite{deligiannis2018secrecy}, where MIMO radar transmits two different signals, carrying desired information and false information, respectively, both of which are employed for sensing. Optimization problems were designed to maximize the secrecy rate for safeguarding the communication data. As studied in \cite{chu2021aided,su2020secure}, the dual-functional base station (BS) detects targets and transmits information to CUs simultaneously, where each of the targets is regarded as a potential eavesdropper. In this scenario, the artificial noise (AN)-aided secure beamforming design enables the secure information transmission from the BS to CUs in ISAC systems. Specifically, AN is generated at the transmitter side to deteriorate the received signal at each target/Eve, thus the decoding capability of which is destructed. To avoid the redundant power consumption caused by the added AN, the research in \cite{su2022secure} proposed a symbol-level precoding algorithm to exploit constructive interference (CI) to aid detection from the legitimate users, and destructive interference (DI) to inhibit detection from the target/Eve. More recently, the encryption keys mechanism has been applied in PLS, where the filterband-based PLS algorithm was proposed to enable key generation by decomposing the received signal in parallel sub-bands, namely chirp modulation \cite{dwivedi2020secure} . This method secured ISAC systems via improving the secret key generation rate efficiently, which however depends on the radio channel charateristics. Additionally, the information-theoretic study in \cite{gunlu2022secure} considered to mitigate information leakage between sensing and communication operations in the ISAC system, where the inner and outer bounds for the secrecy-distortion region were derived under the assumption of perfect and partial output feedback.
\subsection{Contributions}
We note that in the above works on secure ISAC transmission, the radar and communication systems work individually over separate end-goals rather than cooperating with each other. For further promoting the integration of S\&C functionalities to improve the security of the ISAC systems, we propose a novel approach to ensure the PLS for communication data transmission, which is assisted by the sensing functionality. At the first stage, the dual-functional access point (AP) emits an omni-directional waveform for Eve detection, which then receives echoes reflected from both CUs and Eves located within the sensing range. Suppose that all CUs are cooperative users. That is, the location information of each is acknowledged to the AP. Thus, it is possible to obtain angle estimates of Eves contained in the reflected echo by removing known CUs' angles. The estimation performance is measured by the Cram\'{e}r-Rao Bound (CRB) \cite{kay1993fundamentals}.

In the next stage, we formulate a weighted optimization problem to minimize the CRB of targets/Eves and maximize the secrecy rate, subject to beampatern constraints as well as a transmit power budget. A key novelty in this setup is that the channel information in the secrecy rates, is a function of the sensing performance. Specifically, to avoid any false dismissal detection, the main lobe of the beampattern is designed to be wide, with a width depending on the estimation accuracy. Afterwards, by improving estimation accuracy, the sensing and security functionalities provide mutual benefits, resulting in improvement of the mutual performances with every iteration of the optimization, until convergence.
\\\indent Within this scope, the contributions of our work are summarized as follows:
\begin{itemize}
  \item We present a sensing-assisted PLS algorithm of the ISAC system, where the CRB and secrecy rate are employed to measure the sensing and secrecy performance, respectively. Specifically, the secrecy rate is updated with the increasing accuracy of the Eve angle estimation iteratively.
  \item We analyze the lower bound of CRB and the upper bound of the secrecy rate with the constraint of power budget in our proposed ISAC system.
  \item We iteratively maximize the determinant of the Fisher Information Matrix (FIM) and the secrecy rate of the ISAC system by jointly designing the beamforming matrix and the AN.
  \item We further consider the Eve location uncertainty, where the main beam of the sensing beampattern is designed to be sufficiently wide to illuminate the possible angular region that a Eve may appear with high probability, which is indicated by the CRB value obtained from the previous iteration. This implicates that secrecy needs to be provided throughout the angle range.
  \item We design a fractional programming (FP) algorithm to solve the proposed weighted optimizaion problem and verify the efficiency of the solver for both single-Eve and multi-Eve detection.
\end{itemize}
\subsection{Organization}
This paper is organized as follows. Section II gives the system model. Benchmark schemes including AN design techniques with unknown and statistically known Eve channel information are given in Section III. Section IV presents the approach to estimate Eves' parameters. Bounds for the metrics CRB and secrecy rate are given in Section V and the weighted optimization problem is accordingly designed for Eves' parameters estimation and communication data security in Section VI. Section VII provides numerical results, and Section VIII concludes the paper.
\\\indent \emph{Notations}: Unless otherwise specified, matrices are denoted by bold uppercase letters (i.e., $\mathbf{X}$), vectors are represented by bold lowercase letters (i.e., $\mathbf{x}$), and scalars are denoted by normal font (i.e., $\alpha$). Subscripts indicate the location of the entry in the matrices or vectors (i.e., $s_{i,j}$ and $l_n$ are the $(i,j)$-th and the \emph{n}-th element in $\mathbf{S}$ and $\mathbf{l}$, respectively). $\operatorname{tr}\left(\cdot\right)$ and $\operatorname{vec}\left(\cdot\right)$ denote the trace and the vectorization operations. $\left(\cdot\right)^T$, $\left(\cdot\right)^H$ and $\left(\cdot\right)^*$ stand for transpose, Hermitian transpose and complex conjugate of the matrices, respectively. $\operatorname{diag}\left(\cdot\right)$ represents the vector formed by the diagonal elements of the matrices and ${\text{rank}}\left(  \cdot  \right)$ is rank operation. $\left\| \cdot\right\|$, $\left\| \cdot\right\|_{\infty}$ and $\left\| \cdot\right\|_F$ denote the $l_2$ norm, infinite norm and the Frobenius norm respectively. $\mathbb{E}\left\{ \cdot  \right\}$ denotes the statistical expectation.


\begin{figure}
  \centering
  \subfigure[Stage 1--The ISAC AP emits omnibeampattern for Eve estimations]{
  \begin{minipage}[b]{0.35\textwidth}
  \includegraphics[width=1\textwidth]{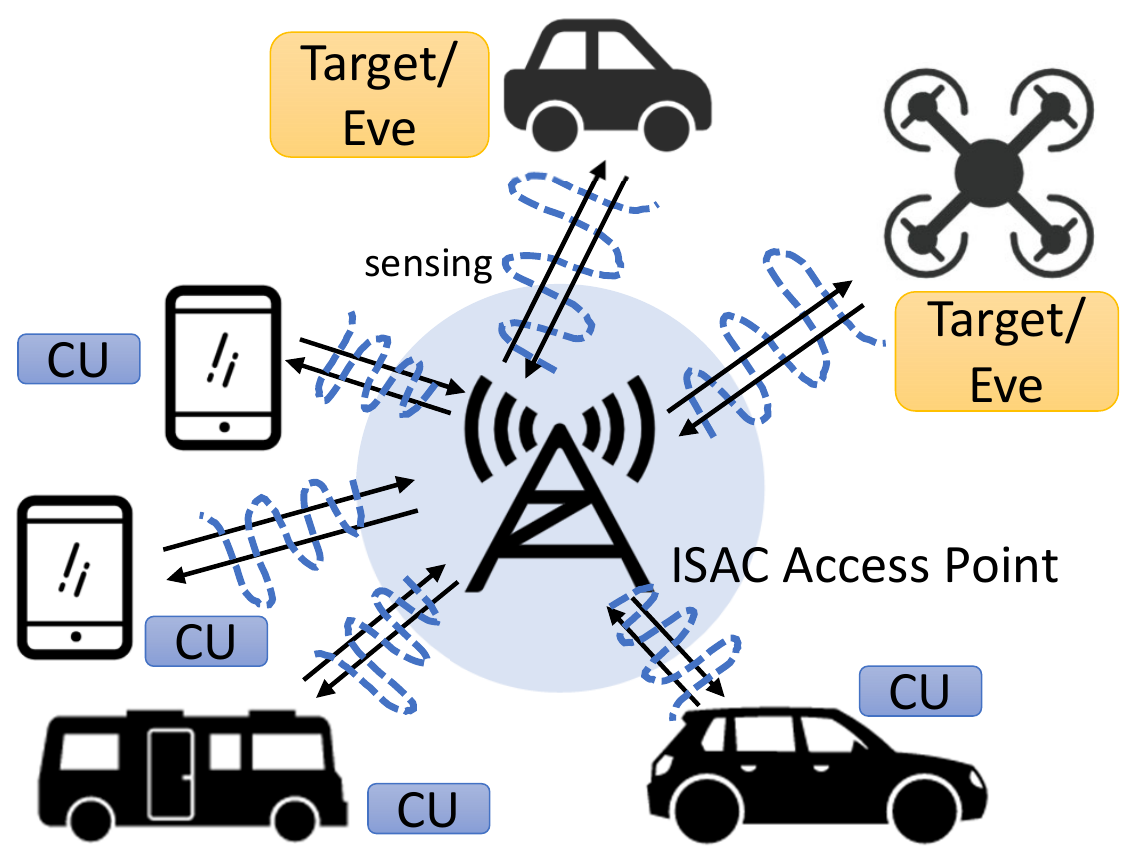}
  \end{minipage}
  }
  \subfigure[Stage 2--Sensing-aided secure ISAC system]{
    \begin{minipage}[b]{0.35\textwidth}
    \includegraphics[width=1\textwidth]{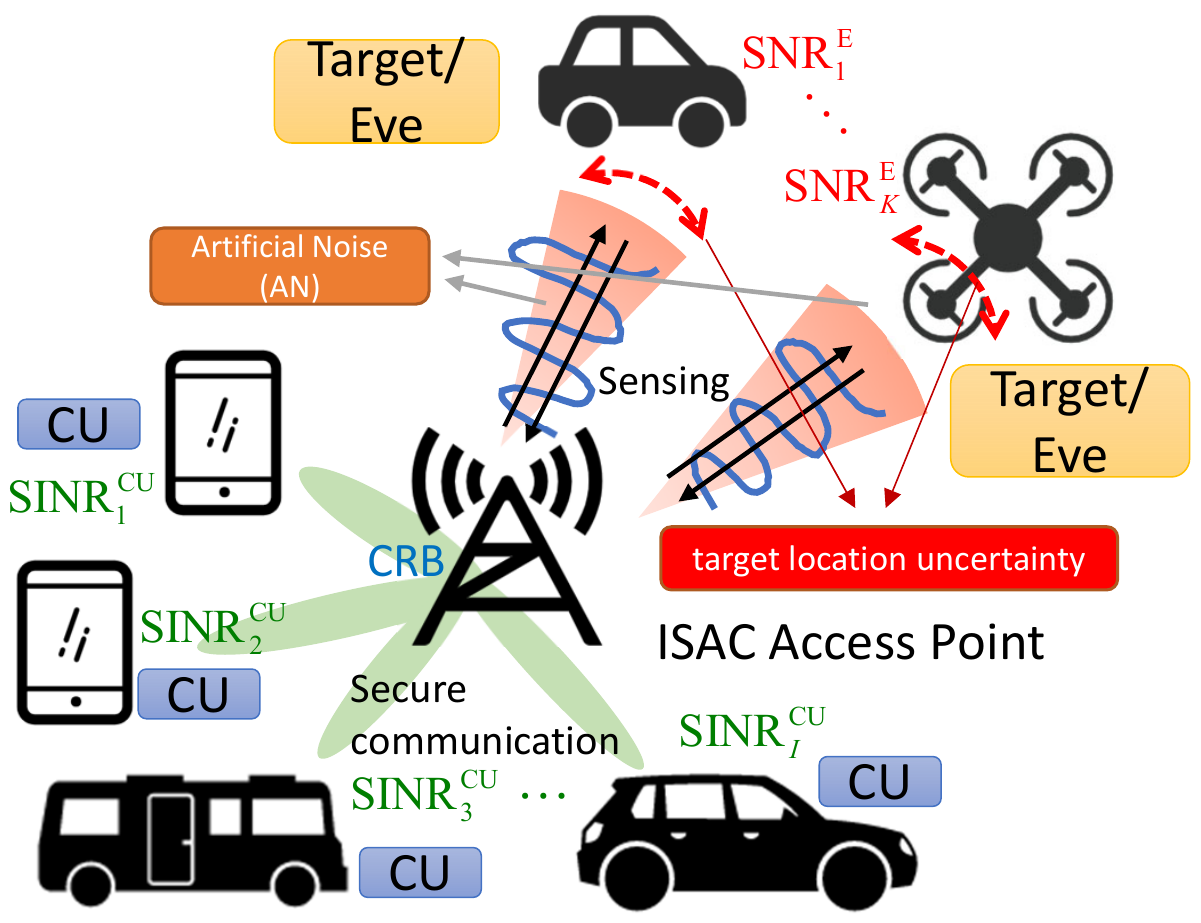}
    \end{minipage}
    }
    \captionsetup{font={footnotesize}}
    \caption{Architecture of the proposed secure ISAC system assisted by the sensing functionality.} \label{fig:1}
\end{figure}

\section{System Model}
We consider a mmWave ISAC system equipped with co-located antennas and let $N_t$ and $N_r$ denote the number of transmit antennas and receive antennas, where the base station communicates with $I$ communication users (CUs) and detects $K$ targets/Eves simultaneously as depicted in Fig. 1. We assume the BS has knowledge of the CUs and their channels, and has no knowledge of the Eves.
\subsection{Communication Signal Model and Metrics}
Let the rows of ${\mathbf{X}} \in {\mathbb{C}^{{N_t} \times L}}$ denote the transmit waveforms, where $L$ is the number of time-domain snapshots. By transmitting the dual-functional waveforms to $I$ CUs, the received signal matrix at the receivers can be expressed as
\begin{equation}\label{eq1}
    \mathbf{Y}_C = \mathbf{HX} + {\mathbf{Z}_C},
\end{equation}
where $\mathbf{Z}_C \in \mathbb{C} ^{I\times L}$ is the additive white Gaussian noise (AWGN) matrix and with the variance of each entry being $\sigma _C^2$. ${\mathbf{H}} = \left[ {{{\mathbf{h}}_1},{{\mathbf{h}}_2}, \ldots ,{{\mathbf{h}}_I}} \right]^H \in \mathbb{C}^{I \times {N_t}}$ represents the communication channel matrix, which is assumed to be known to the BS, with each entry being independently distributed. Following the typical mmWave channel model in \cite{zhao2017tone,su2022secure}, we assume that $\mathbf{h}_i$ is a slow-fading block Rician fading channel. The channel vector of the $i$-th user can be expressed as
\begin{equation}\label{eq2}
    {{\mathbf{h}}_i} = \sqrt {\frac{{{v_i}}}{{1 + {v_i}}}} {\mathbf{h}}_{L,i}^{{\text{LoS}}} + \sqrt {\frac{1}{{1 + {v_i}}}} {\mathbf{h}}_{S,i}^{{\text{NLoS}}},
\end{equation}
where $v_i>0$ is the Rician $K$-factor of the $i$-th user, ${\mathbf{h}}_{L,i}^{{\text{LoS}}} = \sqrt {{{N_t}}} {\mathbf{a}_t}\left( {{\omega _{i,0}}} \right)$ is the LoS deterministic component. ${\mathbf{a}}\left( {{\omega _{i,0}}}\right)$ denotes the array steering vector, where ${\omega _{i,0}} \in \left[ {{\text{ - }}\frac{\pi }{2}, \frac{\pi }{2}} \right]$ is the angle of departure (AOD) of the LoS component from the BS to the user $i$ \cite{hu2019cluster, zhao2017tone}. The scattering component ${\mathbf{h}}_{S,k}^{{\text{NLoS}}}$ can be expressed as ${\mathbf{h}}_{S,i}^{{\text{NLoS}}} = \sqrt {\frac{{{N_t}}}{{{L_p}}}} \sum\nolimits_{l = 1}^{{L_p}} {{c_{i,l}}{{\mathbf{a}}_t}\left( {{\omega _{i,l}}} \right)} $, where $L_p$ denotes the number of propagation paths, ${c_{i,l}} \sim \mathcal{C}\mathcal{N}\left( {0,1} \right)$ is the complex path gain and ${\omega _{i,l}} \in \left[ {{\text{ - }}\frac{\pi }{2}, \frac{\pi }{2}} \right]$ is the AOD associated to the $\left( i,l \right)$-th propagation path.
\\\indent The waveform $\mathbf{X}$ in (\ref{eq1}) can be expressed as
\begin{equation}\label{eq3}
    \mathbf{X} = \mathbf{W} \mathbf{S}+\mathbf{N},
\end{equation}
where $\mathbf{W} \in {\mathbb{C}^{{N_t} \times I}}$ is the dual-functional beamforming matrix to be designed, each row of $\mathbf{S}\in {\mathbb{C}^{{I} \times L}}$ denotes the $i$-th unit-power data stream intended to CUs, and $\mathbf{N} \in {\mathbb{C}^{{N_t} \times L}}$ is the AN matrix generated by the transmitter to interfere potential eavesdroppers. We assume that ${\mathbf{N}} \sim \mathcal{C}\mathcal{N}\left( {{\mathbf{0}},{{\mathbf{R}}_N}} \right)$, where ${{\mathbf{R}}_N} \succeq {\mathbf{0}}$ denotes the covariance matrix of the AN that is to be designed. We further assume that the data streams are approximately orthogonal to each other, yielding
\begin{equation}\label{eq4}
    \frac{1}{L}{{\mathbf{S}}_C}{\mathbf{S}}_C^H \approx {{\mathbf{I}}_{I \times I}}.
\end{equation}
Note that (4) is asymptotically achievable when $L$ is sufficiently large. Then, we denote the beamforming matrix as $\mathbf{W} = \left[ {{{\mathbf{w}}_1}, \ldots ,{{\mathbf{w}}_I}} \right]$, where each column $\mathbf{w}_i$ is the beamformer for the $i$-th CU. Accordingly, the SINR of the $i$-th user is given as
\begin{equation}\label{eq5}
\begin{aligned}
    {\text{SIN}}{{\text{R}}^\text{CU}_i} &= \frac{{{{\left| {{\mathbf{h}}_i^H{{\mathbf{w}}_i}} \right|}^2}}}{{\sum\limits_{m = 1,m \ne i}^I {{{\left| {{\mathbf{h}}_i^H{{\mathbf{w}}_m}} \right|}^2} { + \left| {{\mathbf{h}}_i^H{{\mathbf{R}}_N}{{\mathbf{h}}_i}} \right|}+ \sigma _C^2} }}\\
    &= \frac{{{\text{tr}}\left( {{{{\mathbf{\tilde H}}}_i}{{{\mathbf{\tilde W}}}_i}} \right)}}{{\sum\limits_{m = 1,m \ne i}^I {{\text{tr}}\left( {{{{\mathbf{\tilde H}}}_i}{{{\mathbf{\tilde W}}}_m}} \right) { + {\mathbf{tr}}\left( {{{{\mathbf{\tilde H}}}_i}{{\mathbf{R}}_N}} \right)} + \sigma _C^2} }},
\end{aligned}
\end{equation}
where we denote ${{\mathbf{\tilde H}_i}} = {\mathbf{h}_i}{\mathbf{h}_i^H}$ and ${{\mathbf{\tilde W}_i}} = {\mathbf{w}_i}{\mathbf{w}_i^H}$.
\subsection{Radar Signal Model}
By emitting the waveform $\mathbf{X}$ to sense Eves, the reflected echo signal matrix at the BS receive array is given as
\begin{equation}\label{eq6}
    {\mathbf{Y}}_R = \sum\limits_{k = 1}^K {{{\mathbf{a}}^*}\left( {{\theta _k}} \right){\beta _k}{{\mathbf{b}}^T}} \left( {{\theta _k}} \right){\mathbf{X}} + {\mathbf{Z}_R},
\end{equation}
where ${\mathbf{a}}\left( \theta  \right) \in {\mathbb{C}^{{N_r} \times 1}}$ and ${\mathbf{b}}\left( \theta  \right) \in {\mathbb{C}^{{N_t} \times 1}}$ represent the steering vectors for the receive and transmit arrays, which are assumed to be a uniform linear array (ULA) with half-wavelength antenna spacing. $\beta_k$ is the complex amplitude of the $k$-th Eve. We assume the number of antennas is even and define the receive steering vector as
\begin{equation}\label{eq7}
  {\mathbf{a}}\left( \theta  \right) = {\left[ {{e^{ - j\frac{{{N_r} - 1}}{2}\pi \sin \theta }},{e^{ - j\frac{{{N_r} - 3}}{2}\pi \sin \theta }}, \cdots ,{e^{j\frac{{{N_r} - 1}}{2}\pi \sin \theta }}} \right]^T}.
\end{equation}
It is noted that we choose the center of the ULA antennas as the reference point. To this end, it is easy to verify that
\begin{equation}\label{eq8}
  {{\mathbf{a}}^H}\left( \theta  \right){\mathbf{\dot a}}\left( \theta  \right) = 0.
\end{equation}
Finally, ${\mathbf{Z}_R}$ denotes the interference and the AWGN term. We assume that the columns of ${\mathbf{Z}_R}$ are independent and identically distributed circularly symmetric complex Gaussian random vectors with mean zero and a covariance matrix $\mathbf{Q}=\sigma _R^2{\mathbf{I}}$.
\\\indent Similar to the expression in (5), the eavesdropping SNR received at the $k$-th Eve is written as
\begin{equation}\label{eq9}
    {\text{SN}}{{\text{R}}^\text{E}_k} = \frac{{{{\left| {{\alpha _k}} \right|}^2}{{\mathbf{a}}^H}\left( {{\theta _k}} \right)\sum\limits_{i = 1}^I {{{{\mathbf{\tilde W}}}_i}} {{\mathbf{a}} }\left( {{\theta _k}} \right)}}{{{{\left| {{\alpha _k}} \right|}^2}{{\mathbf{a}}^H}\left( {{\theta _k}} \right){{\mathbf{R}}_N}{\mathbf{a}}\left( {{\theta _k}} \right) + }{\sigma _0^2}},
\end{equation}
where ${\sigma _0^2}$ denotes the covariance of AWGN received by each Eve.
\\\indent For simplicity, the reflected echo signal given in (\ref{eq5}) can be recast as
\begin{equation}\label{eq10}
    {\mathbf{Y}} = {{\mathbf{A}}^*}\left( {\bm{\theta }} \right){\mathbf{\Lambda }}{{\mathbf{B}}^T}\left( {\bm{\theta }} \right){\mathbf{X} } + {\mathbf{Z}_R},
\end{equation}
where we denote ${\mathbf{A}}\left( {\bm{\theta }} \right) = \left[ {{\mathbf{a}}\left( {{\theta _1}} \right), \ldots ,{\mathbf{a}}\left( {{\theta _K}} \right)} \right]$, ${\mathbf{B}}\left( {\bm{\theta }} \right) = \left[ {{\mathbf{b}}\left( {{\theta _1}} \right), \ldots ,{\mathbf{b}}\left( {{\theta _K}} \right)} \right]$, and  ${\mathbf{\Lambda }} = {\text{diag}}\left( {{\beta _k}} \right)$.
\subsection{CRB and Secrecy Rate}
In this subsection, we elaborate on the radar detection and communication security metrics. Particularly, the target/Eve estimation is measured by the CRB, which is a lower bound on the variance of unbiased estimators \cite{9652071}, and the security performance is evaluated by the secrecy rate.
\\\indent In the multi-Eve detection scenario, the CRB with respect to the unknown Eve parameters ${{\theta _1}, \ldots ,{\theta _K}}$ and ${{\beta _1}, \ldots ,{\beta _K}}$ was derived in \cite{li2007range} in detail, and the Fisher information matrix (FIM) for $\theta_k, \forall\;k$ as well as real and imaginary parts of $\beta_k, \forall\;k$ is given as
\begin{equation}\label{eq11}
  {\mathbf{J}} = 2L\left[ {\begin{array}{*{20}{c}}
    {\operatorname{Re} \left( {{{\mathbf{J}}_{11}}} \right)}&{\operatorname{Re} \left( {{{\mathbf{J}}_{12}}} \right)}&{ - \operatorname{Im} \left( {{{\mathbf{J}}_{12}}} \right)} \\
    {{{\operatorname{Re} }^T}\left( {{{\mathbf{J}}_{12}}} \right)}&{\operatorname{Re} \left( {{{\mathbf{J}}_{22}}} \right)}&{ - \operatorname{Im} \left( {{{\mathbf{J}}_{22}}} \right)} \\
    { - {{\operatorname{Im} }^T}\left( {{{\mathbf{J}}_{12}}} \right)}&{ - {{\operatorname{Im} }^T}\left( {{{\mathbf{J}}_{22}}} \right)}&{\operatorname{Re} \left( {{{\mathbf{J}}_{22}}} \right)}
  \end{array}} \right],
\end{equation}
where the elements of the matrix in (11) are given in (12) at the top of next page with $\odot$ denoting the Hadamard (element-wise) matrix product, and ${\mathbf{\dot A}} = \left[ {\begin{array}{*{20}{c}}, {\frac{{\partial {\mathbf{a}}\left( {{\theta _1}} \right)}}{{\partial {\theta _1}}}}&{\frac{{\partial {\mathbf{a}}\left( {{\theta _2}} \right)}}{{\partial {\theta _2}}}}& \ldots &{\frac{{\partial {\mathbf{a}}\left( {{\theta _K}} \right)}}{{\partial {\theta _K}}}} \end{array}} \right]$, ${\mathbf{\dot B}} = \left[ {\begin{array}{*{20}{c}}{\frac{{\partial {\mathbf{b}}\left( {{\theta _1}} \right)}}{{\partial {\theta _1}}}}&{\frac{{\partial {\mathbf{b}}\left( {{\theta _2}} \right)}}{{\partial {\theta _2}}}}& \ldots &{\frac{{\partial {\mathbf{b}}\left( {{\theta _K}} \right)}}{{\partial {\theta _K}}}} \end{array}} \right]$.
\begin{figure*}[tbp]\label{eq12}
  \begin{subequations}
    \begin{align}
     {{\mathbf{J}}_{11}} = &\left( {{{{\mathbf{\dot A}}}^H}{{\mathbf{Q}}^{ - 1}}{\mathbf{\dot A}}} \right) \odot \left( {{{\mathbf{\Lambda }}^ * }{{\mathbf{B}}^H}{\mathbf{R}}_X^ * {\mathbf{B\Lambda }}} \right) + \left( {{{{\mathbf{\dot A}}}^H}{{\mathbf{Q}}^{ - 1}}{\mathbf{A}}} \right) \odot \left( {{{\mathbf{\Lambda }}^ * }{{\mathbf{B}}^H}{\mathbf{R}}_X^ * {\mathbf{\dot B\Lambda }}} \right)
      {\text{        }} + \left( {{{\mathbf{A}}^H}{{\mathbf{Q}}^{ - 1}}{\mathbf{\dot A}}} \right) \odot \left( {{{\mathbf{\Lambda }}^ * }{{{\mathbf{\dot B}}}^H}{\mathbf{R}}_X^ * {\mathbf{B\Lambda }}} \right) + \nonumber \\
      &\left( {{{\mathbf{A}}^H}{{\mathbf{Q}}^{ - 1}}{\mathbf{A}}} \right) \odot \left( {{{\mathbf{\Lambda }}^ * }{{{\mathbf{\dot B}}}^H}{\mathbf{R}}_X^ * {\mathbf{\dot B\Lambda }}} \right) \\
      {{\mathbf{J}}_{12}} = &\left( {{{{\mathbf{\dot A}}}^H}{{\mathbf{Q}}^{ - 1}}{\mathbf{A}}} \right) \odot \left( {{{\mathbf{\Lambda }}^ * }{{\mathbf{B}}^H}{\mathbf{R}}_X^ * {\mathbf{B}}} \right) + \left( {{{\mathbf{A}}^H}{{\mathbf{Q}}^{ - 1}}{\mathbf{A}}} \right) \odot \left( {{{\mathbf{\Lambda }}^ * }{{{\mathbf{\dot B}}}^H}{\mathbf{R}}_X^ * {\mathbf{B}}} \right)  \\
      {{\mathbf{J}}_{22}} = &\left( {{{\mathbf{A}}^H}{{\mathbf{Q}}^{ - 1}}{\mathbf{A}}} \right) \odot \left( {{{\mathbf{B}}^H}{\mathbf{R}}_X^ * {\mathbf{B}}} \right)
    \end{align}
  \end{subequations}
  \rule[-10pt]{18.5cm}{0.05em}
\end{figure*}
Also, the covariance matrix ${\mathbf{R}_X}$ is given as
\begin{equation}\label{eq13}
  \begin{aligned}
  {{\mathbf{R}}_X} &= \frac{1}{L}{\mathbf{X}}{{\mathbf{X}}^H} = {\mathbf{W}}{{\mathbf{W}}^H} +{\mathbf{R}_N} \\ &= \sum\limits_{i = 1}^I {{{{\mathbf{\tilde W}}}_i}}+{\mathbf{R}_N}.
  \end{aligned}
\end{equation}
As per the above, the corresponding CRB matrix is expressed as
\begin{equation}\label{eq14}
  {\text{CRB}}\left( {{\bm{\theta }},{\bm{\beta }}} \right) =  {{{\mathbf{J}}^{ - 1}}}
\end{equation}
and
\begin{equation}\label{eq15}
  \begin{aligned}
  &{\text{CRB}}\left( {\bm{\theta }} \right) = {\left[ {{{\mathbf{J}}^{ - 1}}} \right]_{11}}  \\
  &{\text{CRB}}\left( {\bm{\beta }} \right) = {\left[ {{{\mathbf{J}}^{ - 1}}} \right]_{22}} + {\left[ {{{\mathbf{J}}^{ - 1}}} \right]_{33}}
  \end{aligned}
\end{equation}
\\\indent Moreover, the achievable secrecy rate at the legitimate user is defined as the difference between the achievable rates at the legitimate receivers and the eavesdroppers. Thus, we give the expression of the worst-case secrecy rate as \cite{6820768,su2020secure}
\begin{equation}\label{eq16}
  {\text{SR}} \left( {{{\mathbf{\tilde W}}}_i},{\mathbf{R}_N} \right) = \mathop {\min }\limits_{i,k} {\left[ {{R^\text{CU}_{i}} - {R^\text{E}_{k}}} \right]^ + },
\end{equation}
where ${R^\text{CU}_{i}}, \forall\;i$ and ${R^\text{E}_{k}}, \forall \;k$ represent the achievable transmission rate of the $i$-th CU and the $k$-th Eve, which can be expressed as (14a) and (14b), respectively.
\begin{subequations}\label{eq17}
  \begin{align}
    {R^\text{CU}_{i}} \left( {{{\mathbf{\tilde W}}}_i},{\mathbf{R}_N} \right) = \log \left( {1 + {\text{SIN}}{{\text{R}}^\text{CU}_i}} \right) \\
    {R^\text{E}_{k}} \left( {{{\mathbf{\tilde W}}}_i},{\mathbf{R}_N} \right) = \log \left( {1 + {\text{SN}}{{\text{R}}^\text{E}_k}} \right)
  \end{align}
\end{subequations}

\section{Benchmark Schemes: Isotropic AN-Aided Secure Beamforming and Eve-award AN Design}
In the scenario considered with no knowledge of the Eves, a typical method to avoid the information inception is to transmit AN. To be specific, partial transmit power is allocated to emit the AN to interfere with the Eves, where the AN is isotropically distributed on the orthogonal complement subsapce of CUs' channels \cite{hassibi2002multiple}. To elaborate on this, we firstly take the $l$-th snapshot as a reference, i.e., (1) is simplified as
\begin{equation}\label{eq18}
  {{\mathbf{y}}_C}\left[ l \right] = {\mathbf{Hx}}\left[ l \right] + {{\mathbf{z}}_C}\left[ l \right].
\end{equation}
where ${\mathbf{x}}\left[ l \right] = {\mathbf{Ws}}\left[ l \right] + {\mathbf{n}}\left[ l \right]$. For simplicity, the snapshot index $l$ will be omitted in the following descriptions. We further rewrite the AN vector $\mathbf{n}$ as
\begin{equation}\label{eq19}
  {\mathbf{n}}{\text{ = }}{\mathbf{V\bar n}},
\end{equation}
where $\mathbf{V} = {\mathbf{P}}_{\mathbf{H}}^ \bot  = {{\mathbf{I}}_{{N_t}}} - {\mathbf{H}^H}{\left[ {{{\mathbf{H}}}{\mathbf{H}^H}} \right]^{ - 1}}{{\mathbf{H}}}$ denotes the orthogonal complement projector of the $\mathbf{H}$, and $\mathbf{\bar n}$ is the zero-mean colored noise vector with a covariance matrix ${{\mathbf{R}}_{\bar n}} = \mathbb{E}\left\{ {{\mathbf{\bar n}}{{{\mathbf{\bar n}}}^H}} \right\}$ \cite{liao2010qos, fang2015precoding}. Accordingly, the covariance matrix is given as
\begin{equation}\label{eq20}
  {{{\mathbf{\bar R}}}_x} = \sum\limits_{i = 1}^I {{{{\mathbf{\tilde W}}}_i}}  + {\mathbf{V}}{{\mathbf{R}}_{\bar n}}{{\mathbf{V}}^H}.
\end{equation}
Then, the received signal vector of legitimate CUs is written as
\begin{equation}\label{eq21}
  {{\mathbf{y}}_C} = {\mathbf{HWs}} + {{\mathbf{z}}_C}.
\end{equation}
It is noted that the AN does not interfere the CUs' channels and the SINR of the $i$-th user is given as
\begin{equation}\label{eq22}
  \overline {{\text{SINR}}} _i^{{\text{CU}}} = \frac{{{\text{tr}}\left( {{{{\mathbf{\tilde H}}}_i}{{{\mathbf{\tilde W}}}_i}} \right)}}{{\sum\limits_{m = 1,m \ne i}^I {{\text{tr}}\left( {{{{\mathbf{\tilde H}}}_i}{{{\mathbf{\tilde W}}}_m}} \right)  + \sigma _C^2} }}.
\end{equation}
Likewise, the SNR of the $k$-th Eve is given as
\begin{equation}\label{eq23}
  \begin{aligned}
    \overline {{\text{SNR}}} _k^{\text{E}} &= \frac{{\mathbb{E}\left\{ {{\mathbf{g}}_k^H{\mathbf{Ws}}} \right\}}}{{\mathbb{E}\left\{ {{\mathbf{g}}_k^H{\mathbf{n}}} \right\} + \sigma _0^2}} = \frac{{{\mathbf{g}}_k^H\sum\limits_{i = 1}^I {{{{\mathbf{\tilde W}}}_i}} {{\mathbf{g}}_k}}}{{{\mathbf{g}}_k^H{\mathbf{V}}{{\mathbf{R}}_{\bar n}}{{\mathbf{V}}^H}{{\mathbf{g}}_k} + \sigma _0^2}}\\
    &= \frac{{{\text{tr}}\left( {{{\mathbf{G}}_k}\sum\limits_{i = 1}^I {{{{\mathbf{\tilde W}}}_i}} } \right)}}{{{\text{tr}}\left( {{{\mathbf{G}}_k}{\mathbf{V}}{{\mathbf{R}}_{\bar n}}{{\mathbf{V}}^H}} \right) + \sigma _0^2}},
  \end{aligned}
\end{equation}
where ${{\mathbf{g}}_k}$ denotes the channel from the transmitter to the $k$-th Eve. Note that the covariance matrix of the colored noise vevtor, i.e., ${{\mathbf{R}}_{\bar n}}$, is set as identity matrix when Eves' channels are unknown to the ISAC BS.
\subsection{AN Refinement Based on Eves' Information}
The AN design could be further refined if more information about the Eve's channels $\mathbf{g}_k$ is known to the BS. In this case, we assume that the instantaneous channel realizations of Eves are known to the transmitter, which is defined as ${{\mathbf{G}}_k} = \mathbb{E}\left\{ {{{\mathbf{g}}_k}{\mathbf{g}}_k^H} \right\} = {{{\mathbf{\bar g}}}_k}{\mathbf{\bar g}}_k^H + \sigma _{G,k}^2{{\mathbf{I}}_{{N_t}}}$, where ${{{\mathbf{\bar g}}}_k}$ and $\sigma _{G,k}^2{{\mathbf{I}}_{{N_t}}}$ denote the mean and covariance matrix of ${{{\mathbf{g}}}_k}$, respectively. In particular, to obtain a fair comparison with our approach that assumes no Eves' information, we consider the extreme setting that ${{\mathbf{G}}_k} = \sigma _{g,k}^2{{\mathbf{I}}_{{N_t}}}$, $\sigma _{g,k}^2 > 0$. Besides, we assume that $\mathbf{g}_k$ and $\mathbf{s}$ are independent and identically distributed (i.i.d.). To this end, the expression of the secrecy rate can be accordingly obtained as given in Section II-C, which is written as
\begin{equation}
  {\text{S}}{{\text{R}}_{{\text{IST}}}} = \mathop {\min }\limits_{i,k} {\left[ {\log \left( {1 + {\overline {{\text{SINR}}} _i^{{\text{CU}}}}} \right) - \log \left( {1 + {\overline {{\text{SNR}}} _k^{\text{E}}}} \right)} \right]^ + }.
\end{equation}
In light of the above assumptions, the secrecy rate maximization problem with the omnidirectional beampattern design is given as
\begin{equation}
  \begin{aligned}
    &\mathop {\max }\limits_{{{{\mathbf{\tilde W}}}_i},{{\mathbf{R}}_{\bar n}}}\;\;\; {\text{  S}}{{\text{R}}_{{\text{IST}}}} \hfill \\
    &{\text{s}}{\text{.t}}{\text{.}}\;\;{{{\mathbf{\bar R}}}_X} = \frac{{{P_0}}}{{{N_t}}}{{\mathbf{I}}_{{N_t}}} \hfill \\
    &\;\;\;\;\;\;{{{\mathbf{\tilde W}}}_i}\succeq  {\mathbf{0}},{{\mathbf{R}}_{\bar n}} \succeq  {\mathbf{0}},\forall i. \hfill \\
  \end{aligned}
\end{equation}
Note that the non-convexity of the problem above only lies in the objection function, while it can be regarded as a typical secrey rate maximization problem, which has been solved efficiently as studied in \cite{chu2015secrecy,li2013spatially}. The simulation results will be given in Section VII as benchmarks.
\section{Eves' Parameters Estimation}
To avoid redundancy, we briefly present the method to estimate amplitutes and angles of Eves based on our signal models proposed in Section II, namely the combined Capon and approximate maximum likelihood (CAML) approach \cite{jakobsson2000combining, 4655353}. Specifically, Capon is initially applied to estimate the peak directions, and then AML is used to estimate the amplitudes of all Eves.
\\\indent We firstly give the expression of signal model $\mathbf{Y}$ \cite{1206680}, where we let ${{\hat \theta }_k}, k = 1, \ldots ,K$ denote the estimated Eves' directions. Similar to the reveive signal model in (8), we here have
\begin{equation}\label{eq15}
  {\mathbf{Y}} = {{\mathbf{A}}^*}\left( {\bm{\hat\theta }} \right){\mathbf{\hat \Lambda }}{{\mathbf{B}}^T}\left( {\bm{\hat \theta }} \right){\mathbf{X} } +  {\mathbf{\tilde Z}},
\end{equation}
where ${\bm{\hat\Lambda }} = {\text{diag}}\left[ {\beta \left( {{{\hat \theta }_1}} \right), \ldots ,\beta \left( {{{\hat \theta }_K}} \right)} \right]$ and ${\mathbf{\tilde Z}}$ denotes the residual term. By employing the approximate maximum likelihood (AML) algorithm, the estimate of amplitutes can be written in a closed form given as \cite{4655353}
\begin{equation}\label{eq16}
  \begin{aligned}
  \bm{\beta} {\text{ = }}&\frac{1}{L}{\left[ {\left( {{{\mathbf{A}}^H}{{\mathbf{T}}^{ - 1}}{\mathbf{A}}} \right) \odot \left( {{{\mathbf{B}}^H}{\mathbf{\hat R}}_X^*{\mathbf{B}}} \right)} \right]^{ - 1}} \cdot \hfill \\
  &\;\;\;\;\;\;\;\;\;\;\;\;\;\;\;\;\;\;\;\;\;\;\;{\text{vecd}}\left( {{{\mathbf{A}}^H}{{\mathbf{T}}^{ - 1}}{\mathbf{Y}}{{\mathbf{X}}^H}{{\mathbf{B}}^*}} \right),
  \end{aligned}
\end{equation}
where vecd($\cdot$) denotes a colomn vector with the elements being the diagnal of a matrix and
\begin{equation}\label{eq17}
  {\mathbf{T}} = L{\mathbf{\hat R}} - \frac{1}{L}{\mathbf{Y}{\mathbf{X}^H}}{{\mathbf{B}}^*}{\left( {{{\mathbf{B}}^T}{{{\mathbf{\hat R}}}_X}{{\mathbf{B}}^*}} \right)^{ - 1}}{{\mathbf{B}}^T}{\mathbf{X}}{{\mathbf{Y}}^H},
\end{equation}
where $\mathbf{\hat R}$ is the sample covariance of the observed data samples and ${\mathbf{\hat R}} = \frac{1}{L}{\mathbf{Y}}{{\mathbf{Y}}^H}$.
\\\indent At the first step of the Eve parameter estimation, we design our transmission so that the AP emits an omnidirectional waveform, which is usually employed by the MIMO radar for initial probing. Thus, the covariance matrix is given as ${{\mathbf{\tilde R}}_X} = \frac{P_0}{N_t}{\mathbf{I}_{N_t}}$. The CRBs for angles and amplitudes of tagets can be accordingly calculated by substituting ${{\mathbf{\tilde R}}_X}$ into (12) and (15), where we denote them as ${\text{CR}}{{\text{B}}_0}\left( {\bm{\hat \theta }} \right)$ and ${\text{CR}}{{\text{B}}_0}\left( {\bm{\hat \beta }} \right)$. Assume that the probability density function (PDF) of the angle estimated error is modeled as Gaussian distribution, zero mean and a variance of ${\text{CR}}{{\text{B}}_0}\left( {\bm{\hat \theta }} \right)$. That is, $\text{E}_{est,k}\sim \mathcal{C}\mathcal{N}\left( {0,{\text{CR}}{{\text{B}}_0}\left( {{\hat \theta_k }} \right)} \right) $, where $\text{E}_{est,k}$ denotes the angle estimation error of the $k$-th Eve. As a consequence, the probability that the real direction of the $k$-th Eve falls in the range ${\Xi^{\left( 0 \right)}_k} = \left[ {\hat \theta_k } - 3\sqrt{{\text{CR}}{{\text{B}}_0}\left( {{\hat \theta_k }} \right)}, {\hat \theta_k } + 3\sqrt{{\text{CR}}{{\text{B}}_0}\left( {{\hat \theta_k }} \right)}\right]$ is approximately 0.9973 \cite{chandola2009anomaly}. Thus, the main lobe width of the radar beampattern will be initially designed as $\Xi^{\left( 0 \right)}$, and then it will be iteratively updated based on the optimized CRB.
\\\indent For clarity, we present the spatial spectrum of the direction of angle (DOA) estimation by deploying the CAML technique in Fig. 2. It is assumed that two Eves are located at $\theta_1=-{25^ \circ }, \theta_2={15^ \circ }$ (denoted by blue lines) and three CUs locate at $\theta_3={40^ \circ }, \theta_4={10^ \circ }, \theta_5=-{30^ \circ }$ (denoted by green lines), with the modulus of complex amplitudes $\beta_1 = 1,\beta_2 = 5, \beta_3 = 4,\beta_4 = 5 $ and $\beta_5 = 2$, where directions of CUs are known to the transnmitter. Fig. 2(a) and Fig. 2(b) demonstrate the CAML performance when SNR=20dB and SNR=-15dB, respectively. It is noted that the CAML approach estimates the DOA precisely when SNR is 20dB, while errors of the angle estimation happen when the SNR decreases to -15dB. To further illustrate the performance of CAML estimation method, the root mean square error (RMSE) versus the SNR of the echo signal is shown in Fig. 3 with the CRB as a baseline. As expected, the CRB is shown as the lower bound of the RMSE obtained by CAML estimation, in particular, the CRB gets tight in the high-SNR regime.
\begin{figure}
  \centering
  \subfigure[]{
  \includegraphics[width=0.48\columnwidth]{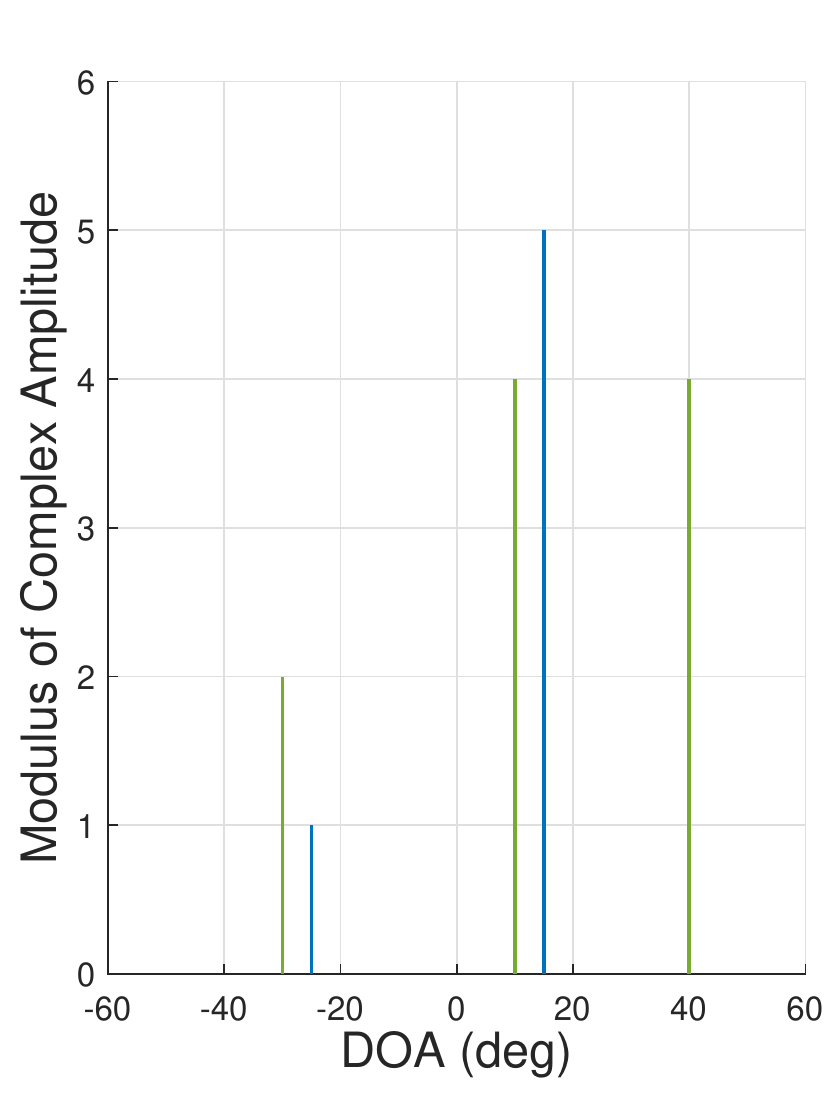}}
  \subfigure[]{
  \includegraphics[width=0.48\columnwidth]{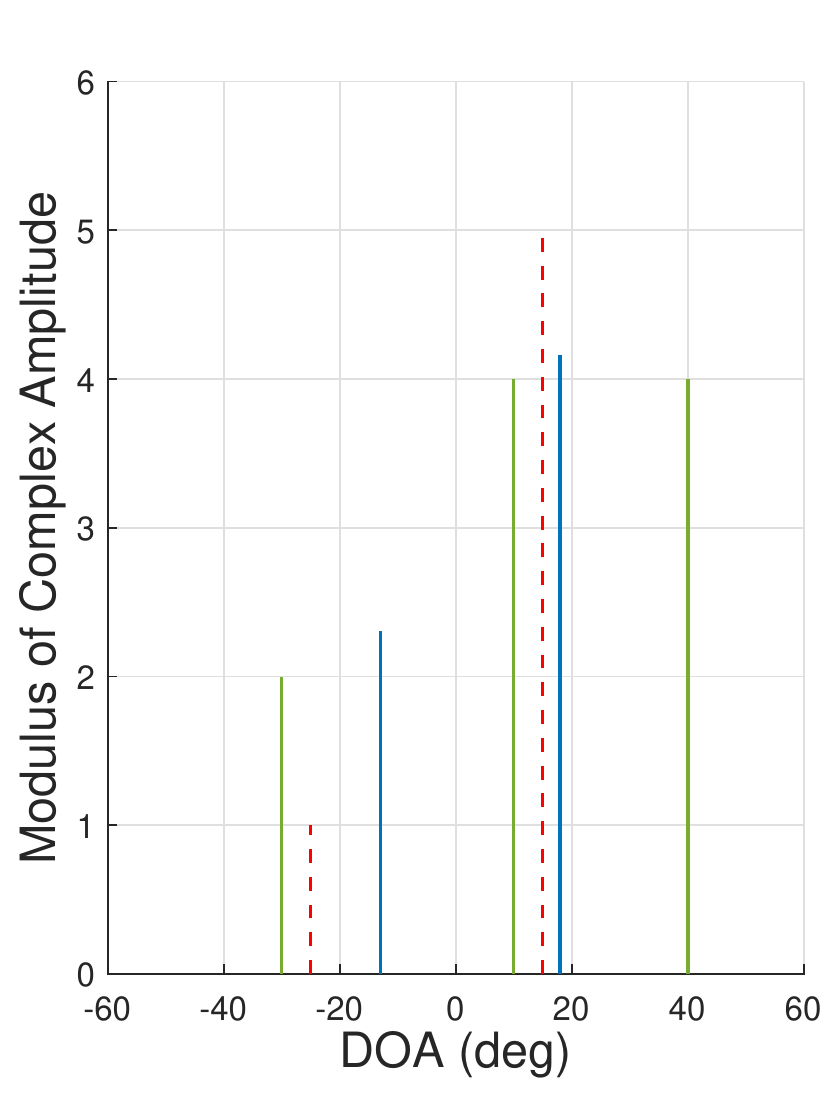}}
  \captionsetup{font={footnotesize}}
  \caption{Spatial spectral estimates with CAML approach, when Eves locate at $\theta_1=-{25^ \circ }, \theta_2={15^ \circ }$ (blue lines), and CUs locate at $\theta_3={40^ \circ }, \theta_4={10^ \circ }$ and $\theta_5=-{30^ \circ }$ (green lines). The red dashed lines in (b) denote the real directions and amplitudes of Eves. (a) SNR=20dB. (b) SNR=-15dB.}
  \label{fig.2}
\end{figure}
\begin{figure}
  \centering
  \includegraphics[width=1\columnwidth]{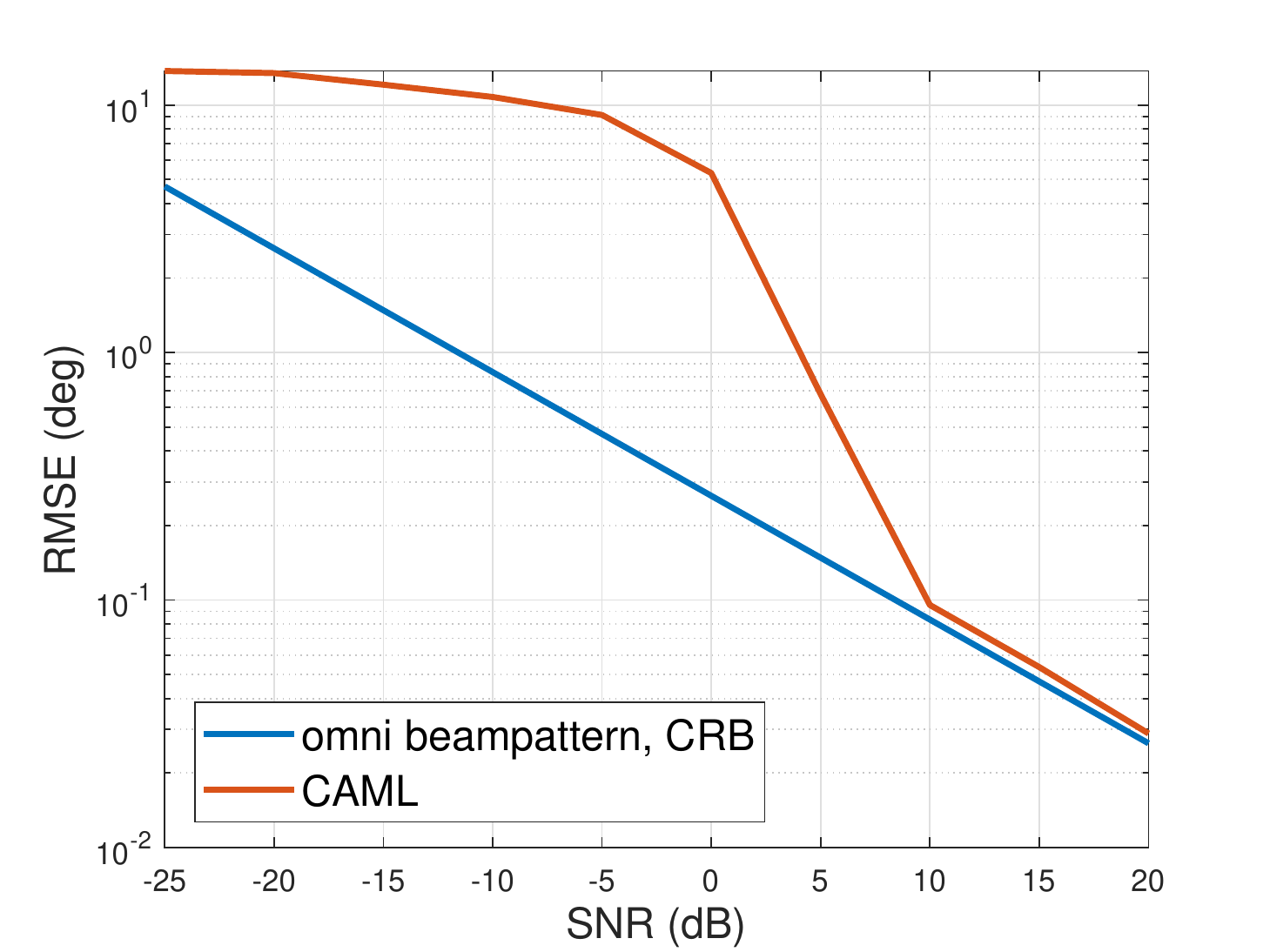}
  \captionsetup{font={footnotesize}}
  \caption{Target/Eve estimation performance by applying CAML method, with the CRB obtained by omni directional beampattern design as a benchmark.}
  \label{fig.3}
\end{figure}
\section{Bounds for CRB and Secrecy Rate}
The design of a wieghted optimization between the radar CRB and the communication secrecy rate presents the challenge that the two perofrmance metrics have different units and potentially different magnitutes. To overcome this challenge we need to normalise them each with their respective upper/lower bound. To obtain these bounds, in this section we present the CRB minimization problem and the secrecy rate maximization problem with the system power budget constraint. Considering the further design of weighted objective function in the following section, the CRB minimization problem can be approximated as the FIM determinant maximization problem. To this end, the optimal solutions generate the upper bounds of the FIM determinant and the secrecy rate, both of which will be employed to normalize the metrics in Section VI.
\subsection{Upper-bound of the FIM Determinant}
We denote $\bm{\eta}$ as the sensing parameters, thus the MSE can be expressed as ${\mathbf{M}}\left( {\bm{\eta }} \right){{ \triangleq }}\mathbb{E}\left\{ \left( {{\bm{\eta }} - {\bm{\hat \eta }}} \right){\left( {{\bm{\eta }} - {\bm{\hat \eta }}} \right)^T} \right\}  \succeq {\mathbf{J}^{-1}} $. For the $m$-th parameter $\eta_m$ to be estimated, it has $\mathbb{E}\left\{ {{{\left\| {{\eta _m} - {{\hat \eta }_m}} \right\|}^2}} \right\} \geq {\left[ {{{\mathbf{J}}^{ - 1}}} \right]_{mm}}$ \cite{tichavsky1995posterior}. Thus, it is commonly to minimize the trace or the determinant of the CRB matrix, i.e., $\text{tr}\left( {\mathbf{J}^{-1}}\right)$ or $\left| {\mathbf{J}^{-1}} \right|$. Since the CRB matrix is the inverse of the FIM matrix, the problem of minimizing $\left| {\mathbf{J}^{-1}}\right|$ is equivalent to maximizing $\left| {\mathbf{J}}\right|$, which is given as \cite{li2007range}
\begin{subequations}\label{eq18}
  \begin{align}
    &\mathop {\max }\limits_{{{{\mathbf{\tilde W}}}_i},{\mathbf{R}_N}} \;\;\; \left| {\mathbf{J}} \right| \hfill \\
    &{\text{s}}{\text{.t}}{\text{.}}\;\; {{{\mathbf{R}}}_N} \succeq  {\mathbf{0}}, {{{\mathbf{\tilde W}}}_i} \succeq {\mathbf{0}}, \forall \; i \hfill \\
    &\;\;\;\;\;\;\;{\text{tr}}\left( {\sum\limits_{i = 1}^I {{{{\mathbf{\tilde W}}}_i}} } + {\mathbf{R}_N}\right) = {P_0},
  \end{align}
\end{subequations}
where $P_0$ denotes the power budget of the proposed system. It is noted that the optimization above is convex and can be efficiently solved by cvx toolbox \cite{grant2014cvx, wu1996software}. Consequently, by substituting the optimal ${{{{\mathbf{\tilde W}}}_i},{\mathbf{R}_N}}$ in (11), the upper-bound of FIM determinant is obtained.
\subsection{Secrecy Rate Bound}
To derive the upper-bound of secrecy rate, we only consider the communication security metric in this subsection. Assuming that the CSI is perfectly known to the BS, the secrecy rate maximization problem can be formulated as
\begin{subequations}\label{eq19}
  \begin{align}
    {\text{SR}^\star} = &\mathop {\max }\limits_{{{{\mathbf{\tilde W}}}_i},{\mathbf{R}_N}} \mathop {\min} \limits_{i,k} {\text{ SR }}\left( {{{\mathbf{\tilde W}}}_i},{\mathbf{R}_N} \right) \hfill \\
    &{\text{s}}{\text{.t}}{\text{.      }}{\text{   }} \left( {29b} \right),(29c).
  \end{align}
\end{subequations}
It is noted that the non-convexity lies in the objective function of (30), which makes the optimization problem above difficult to solve. To resolve this issue, we introduce an auxilary variable $b$, where (30) has the same optimal solutions as the reformulation below
\begin{equation}\label{eq20}
  \begin{aligned}
    {\text{SR}^\star} = &\mathop {\max }\limits_{{{{\mathbf{\tilde W}}}_i},{{\mathbf{R}}_N},b} {\text{ }}\mathop {\min }\limits_{i,k} {\text{ }}\left[ {{R_{{C_i}}} \left( {{{\mathbf{\tilde W}}}_i},{\mathbf{R}_N} \right) - \log b} \right]{\text{ }} \hfill \\
    &{\text{s}}{\text{.t}}{\text{.      }}{\text{   }}\log \left( {1 + \frac{{{{\left| {{\alpha _k}} \right|}^2}{\mathbf{a}^H}\left( {{\theta _k}} \right)\sum\limits_{i = 1}^I {{{{\mathbf{\tilde W}}}_i}} {\mathbf{a} }\left( {{\theta _k}} \right)}}{{{{\left| {{\alpha _k}} \right|}^2}{{\mathbf{a}}^H}\left( {{\theta _k}} \right){{\mathbf{R}}_N}{\mathbf{a}}\left( {{\theta _k}} \right) + \sigma _R^2}}} \right) \\
    &\;\;\;\;\;\;\;\;\;\;\;\;\;\;\;\;\;\;\;\;\;\;\;\;\;\;\;\;\;\;\;\;\;\;\;\;\;\;\;\;\;\;\;\;\;\;\;\;\;\;\;\;\;\; \leq \log b, \forall\; k \hfill \\
    &\;\;\;\;\;\;\;\; \left( {29b} \right),(29c) \hfill \\
  \end{aligned}
\end{equation}
The above problem can be simply relaxed into a convex SDP problem. For brevity, we refer readers to \cite{li2013spatially} for more details.

\section{Weighted Optimization for Eves' Estimation and Secure Communication}
In this section, we propose a normalized weighted optimization problem that reveals the performance tradeoff between the communiaction security and Eve parameters estimation. Additionally, recall that the ISAC access point firstly emits an omnidirectional beampattearn as given in Section IV, where imprecise angles of Eves have been obtained at the given SNR, with the angular uncertainty interval of the $k$-th Eve is denoted as ${\Xi^{\left( 0 \right)}_k}$. To reduce angle estimation errors, we also take the wide main beam design into account, which covers all possible directions of Eves.
\subsection{Problem Formulation}
To achieve the desirable tradeoff between the communication data security and the radar estimation CRB, while taking the estimation errors of Eves' angles and the system power budget into account, we formulate the weighted optimization problem as follows
\begin{subequations}\label{eq28}
  \begin{align}
    &\mathop {\max }\limits_{{{{\mathbf{\tilde W}}}_i},{\mathbf{R}_N}} {\text{  }}\rho \frac{{\left| {\mathbf{J}} \right|}}{{{{\left| {\mathbf{J}} \right|}_{UB}}}} + \left( {1 - \rho } \right)\frac{{\text{SR}}}{{{\text{SR}_{UB}}}} \hfill \\
    &{\text{s}}{\text{.t}}{\text{.      }}{{\mathbf{a}}^H}\left( {{\vartheta _{k,0}}} \right){{\mathbf{R}}_X}{{\mathbf{a}}}\left( {{\vartheta _{k,0}}} \right) - {{\mathbf{a}}^H}\left( {{\vartheta _{k,p}}} \right){{\mathbf{R}}_X}{{\mathbf{a}}}\left( {{\vartheta _{k,p}}} \right) \geq {\gamma _s}, \nonumber \\
    &\;\;\;\;\;\;\;\;\;\;\;\;\;\;\;\;\;\;\;\;\;\;\;\;\;\;\;\;\;\;\;\;\;\;\;\;\;\;\;\;\;\;\;\;\;\;\;\;\;\;\;\;\forall {\vartheta _{k,p}} \in {\text{card}}\left( \Psi _k  \right), \forall\;k  \\
    &\;\;\;\;{\text{      }}{{\mathbf{a}}^H}\left( {{\vartheta _{k,n}}} \right){{\mathbf{R}}_X}{{\mathbf{a}}}\left( {{\vartheta _{k,n}}} \right) \leq \nonumber \\
    &\;\;\;\;\;\;  \left( {1 + \alpha } \right){{\mathbf{a}}^H}\left( {{\vartheta _{k,0}}} \right){{\mathbf{R}}_X}{{\mathbf{a}}}\left( {{\vartheta _{k,0}}} \right),\forall \; {\vartheta _{k,n}} \in {\text{card}}\left( \Omega_k  \right), \forall\;k \hfill \\
    &\;\;\;\;{\text{      }}{{\mathbf{a}}^H}\left( {{\vartheta _{k,n}}} \right){{\mathbf{R}}_X}{{\mathbf{a}}}\left( {{\vartheta _{k,n}}} \right) \geq \nonumber  \\
    &\;\;\;\;\;\; \left( {1 - \alpha } \right){{\mathbf{a}}^H}\left( {{\vartheta _{k,0}}} \right){{\mathbf{R}}_X}{{\mathbf{a}}}\left( {{\vartheta _{k,0}}} \right), \forall \;{\vartheta _{k,n}} \in {\text{card}}\left( \Omega_k  \right), \forall\;k \\
    &\;\;\;\;\; \left( {29b} \right),(29c)
  \end{align}
\end{subequations}
where ${{{{\left| {\mathbf{J}} \right|}_{UB}}}}$ and ${{{\text{SR}_{UB}}}}$ denote the upper bounds of the FIM matrix determinant and the secrecy rate which were obtained in Section IV, respectively. $0 \leq \rho \leq 1$ denotes the weighting factor that determines the weights for the Eve estimation performance and the secrecy rate. $\alpha$ denotes a given scalar associated with the wide main beam fluctuation. $\vartheta_{k,n}$ is the $n$-th possible direction of the $k$-th Eve, $\vartheta_{k,0}$ is the angle which was estimated by the algorithm proposed in Section IV. $\Omega_k$ and $\Phi_k$ denote the main beam region and sidelobe region, respectively. Note that $ {\text{card}}\left( \cdot  \right)$ denotes the the cardinality of $\left( \cdot  \right)$.

\begin{remark}
 It is important to highlight that the secrecy rate given by (16) is a function of the estimation accuracy of the Eve's parameters, including $\theta_k$ and $\alpha_k$. Accordingly, beyond the tradeoff in the weighted optimization in this section, the improvement in the sensing performance directly results in an improvement in the secrecy performance.
\end{remark}
\renewcommand{\algorithmicrequire}{\textbf{Input:}}
\renewcommand{\algorithmicensure}{\textbf{Output:}}
\begin{algorithm}[t!]
\label{alg:1}
  \caption{Iterative optimization of the CRB and the secrecy rate}
  \begin{algorithmic}[1]
  \renewcommand{\algorithmicrequire}{\textbf{Initialization:}}
  \REQUIRE
  ${\Xi^{\left( 0 \right)}_k}$ obtained from initial target/Eve estimation and CRB in Section IV; $r = 1$
  \REPEAT
  \STATE
  $\Omega^{\left(r \right)}_k = {\Xi^{\left( r-1 \right)}_k}$, $\Psi^{\left( r \right)}_k$ is acccordingly obtained;
  \STATE
  substitute $\Omega^{\left( r \right)}_k$ and $\Psi^{\left( r \right)}_k$ into problem (32);
  \REPEAT
  \STATE solve problem (32) by FP algorithm;
  \UNTIL find the optimal $c \in \left[ {{{\left( {\mathop {\min }\limits_i {\text{ }}1 + {P_0}{{\left\| {{{\mathbf{h}}_i}} \right\|}^2}} \right)}^{ - 1}},1} \right]$ which generates the maximum value of the objective function deploying the golden search;
  \STATE
  the optimal variables ${{{{\mathbf{\tilde W}}}^\star_i},{{\mathbf{R}}^\star_N}}$ are obtained;
  \STATE
  calculate the $\text{CRB}_r \left(\hat \theta \right)$ and the secrecy rate in the $r$-th iteration;
  \STATE
  ${\Xi^{\left( r \right)}_k}$ can be acccordingly obtained;
  \STATE
  update $r = r + 1$,
  \UNTIL Convergence.
  \end{algorithmic}
\end{algorithm}
\subsection{Efficient Solver}
To tackle problem (32), we firstly recast the complicated secrecy rate term in the objective function. For simplicity, we denote ${\Sigma _i} = \sum\nolimits_{m = 1}^I {{\text{tr}}\left( {{{{\mathbf{\tilde H}}}_i}{{{\mathbf{\tilde W}}}_m}} \right)} $ and rewrite the optimization problem as (33).
\begin{figure*}\label{eq33}
  \begin{subequations}
  \begin{align}
    &\mathop {\max }\limits_{{{{\mathbf{\tilde W}}}_i},{\mathbf{R}_N}} {\text{  }}\frac{\rho}{{\left| {\mathbf{J}} \right|}_{UB}} \left| {\mathbf{J}} \right| + \frac{1-\rho}{\text{SR}_{UB}} \mathop {\min }\limits_{i,{{k,n}}} {\left[ {R_{{C_i}}} \left( {{{\mathbf{\tilde W}}}_i},{\mathbf{R}_N} \right) - {\log \left( {1 + \frac{{{{\left| {{\alpha _k}} \right|}^2}{{\mathbf{a}}^H}\left( {{\vartheta _{k,n}}} \right)\sum\limits_{i = 1}^I {{{{\mathbf{\tilde W}}}_i}} {{\mathbf{a}} }\left( {{\vartheta _{k,n}}} \right)}}{{{{\left| {{\alpha _k}} \right|}^2}{{\mathbf{a}}^H}\left( {{\vartheta _{k,n}}} \right){{\mathbf{R}}_N}{\mathbf{a}}\left( {{\vartheta _{k,n}}} \right) + }{\sigma _R^2}}} \right)} \right]^ + },\nonumber \hfill \\
    &\;\;\;\;\;\;\;\;\;\;\;\;\;\;\;\;\;\;\;\;\;\;\;\;\;\;\;\;\;\;\;\;\;\;\;\;\;\;\;\;\;\;\;\;\;\;\;\;\;\;\;\;\;\;\;\;\;\;\;\;\;\;\;\;\;\;\;\;\;\;\;\;\;\;\;\;\;\;\;\;\;\;\;\;\;\;\;\;\;\;\;\;\;\;\;\;\;\;\;\;\;\;\;\;\;\;\;\;\;\;\;\;\;\;\;\;\;\;\;\;\;\;\;\;\;\;\;\;\;{\vartheta _{k,n}} \in {\text{card}}\left( \Omega_k  \right), \forall \; k  \hfill \\
    &{\text{s}}{\text{.t}}{\text{.  }}\left( {32b} \right),\left( {32c} \right), \left( {32d} \right) \;\text{and} \; \left( {32e} \right)
  \end{align}
  \end{subequations}
  \rule[-10pt]{18.5cm}{0.05em}
\end{figure*}
According to \cite{li2013spatially}, the weighted optimization problem can be recast as (34) by introducing the scalar $b$. Problem (33) and (34) are shown at the top of next page.
\begin{figure*}\label{eq34}
  \begin{subequations}
  \begin{align}
    &\mathop {\max }\limits_{{{{\mathbf{\tilde W}}}_i},{{\mathbf{R}}_N}} {\text{  }}\mathop {{\text{min}}}\limits_i \left(\frac{\rho }{{{{\left| {\mathbf{J}} \right|}_{UB}}}}\left| {\mathbf{J}} \right| + \frac{{ {1 - \rho } }}{{{2^{S{R_{UB}}}}}}\frac{{\Sigma_i  + {\text{tr}}\left( {{{{\mathbf{\tilde H}}}_i}{{\mathbf{R}}_N}} \right) + 1}}{{b\left( {\Sigma_i - {\text{tr}}\left( {{{{\mathbf{\tilde H}}}_i}{{{\mathbf{\tilde W}}}_i}} \right) + {\text{tr}}\left( {{{{\mathbf{\tilde H}}}_i}{{\mathbf{R}}_N}} \right) + 1} \right)}} \right)\hfill \\
    &{\text{s}}{\text{.t}}{\text{.  }}{\text{   }}\frac{{{{\left| {{\alpha _k}} \right|}^2}{{\mathbf{a}}^H}\left( {{\vartheta _{k,n}}} \right)\sum\limits_{i = 1}^I {{{{\mathbf{\tilde W}}}_i}} {\mathbf{a}}\left( {{\vartheta _{k,n}}} \right)}}{{{{\left| {{\alpha _k}} \right|}^2}{{\mathbf{a}}^H}\left( {{\vartheta _{k,n}}} \right){{\mathbf{R}}_N}{\mathbf{a}}\left( {{\vartheta _{k,n}}} \right) + 1}} \leq b - 1,\forall\; {\vartheta _{k,n}} \in {\text{card}}\left( {{\Omega _k}} \right),\forall \; k \hfill \\
    &\;\;\;\;\;\;\;\left( {32b} \right),\left( {32c} \right), \left( {32d} \right) \;\text{and} \; \left( {32e} \right)
  \end{align}
  \end{subequations}
  \rule[-10pt]{18.5cm}{0.05em}
\end{figure*}
\\\indent It is noted that the min operator only applies to the second term of the objective function of problem (34). According to the Fractional Programming (FP) algorithm \cite{shen2018fractional}, the optimization problem can be further reformulated by replacing the fraction term with the coefficient $z$, which is given as
\begin{subequations}
  \begin{align}
    &\mathop {\max }\limits_{{{{\mathbf{\tilde W}}}_i},{{\mathbf{R}}_N},\mathbf{y}, z} {\text{  }}\frac{\rho }{{{{\left| {\mathbf{J}} \right|}_{UB}}}}\left| {\mathbf{J}} \right| + \frac{ {1 - \rho } }{{{2^{S{R_{UB}}}}}}z \hfill \\
    &{\text{s}}{\text{.t}}{\text{.  }}{\text{   }}{\text{    }}2{y_i}\sqrt {\Sigma_i + {\text{tr}}\left( {{{{\mathbf{\tilde H}}}_i}{{\mathbf{R}}_N}} \right) + 1}  - \nonumber \hfill\\
    &\;\;\;\;\;y_i^2\left( {b\left( {{\Sigma_i} - {\text{tr}}\left( {{{{\mathbf{\tilde H}}}_i}{{{\mathbf{\tilde W}}}_i}} \right) + {\text{tr}}\left( {{{{\mathbf{\tilde H}}}_i}{{\mathbf{R}}_N}} \right) + 1} \right)} \right) \geq z, \nonumber \hfill\\
    &\;\;\;\;\;\;\;\;\;\;\;\;\;\;\;\;\;\;\;\;\;\;\;\;\;\;\;\;\;\;\;\;\;\;\;\;\;\;\;\;\;\;\;\;\;\;\;\;\;\;\;\;\;\;\;\;\;\;\;\;\;\;\;\;\;\;\;\;\;\;\;\;\;\;\;\;\;\;\;\;\;\;\;\;\forall \; i \hfill \\
    &\;\;\;\;\;\;\left( {34b} \right), \left( {32b} \right),\left( {32c} \right), \left( {32d} \right) \;\text{and} \; \left( {32e} \right),
  \end{align}
\end{subequations}
 where $\mathbf{y}$ denotes a collection of variables ${\mathbf{y}} = \left\{ {{y_1}, \ldots ,{y_I}} \right\}$. Referring to \cite{li2013spatially}, let $c = \frac{1}{b}$, where $c \in \left[ {{{\left( {\mathop {\min }\limits_i {\text{ }}1 + {P_0}{{\left\| {{{\mathbf{h}}_i}} \right\|}^2}} \right)}^{ - 1}},1} \right]$. Thus, problem (35) can be rewritten as (37) (next page) by replacing $b$ with $c$, and the optimal $y_i$ can be found in the following closed form
 \begin{equation}
  {y_i} = \frac{{c\sqrt {{\Sigma _i} + {\text{tr}}\left( {{{{\mathbf{\tilde H}}}_i}{{\mathbf{R}}_N}} \right) + 1} }}{{{\Sigma _i} - {\text{tr}}\left( {{{{\mathbf{\tilde H}}}_i}{{{\mathbf{\tilde W}}}_i}} \right) + {\text{tr}}\left( {{{{\mathbf{\tilde H}}}_i}{{\mathbf{R}}_N}} \right) + 1}}.
 \end{equation}
 Note that problem (37) (at the top of next page) can be efficiently solved by the cvx toolbox \cite{grant2014cvx, wu1996software}. Given the interval of $c$, the optimal variables. ${{{{\mathbf{\tilde W}}}^\star_i},{{\mathbf{R}}^\star_N},z^\star}$ can be consequently obtained by performing a one-dimensional line search over $c$, such as uniform sampling or the golden search \cite{bertsekas1997nonlinear}. To this end, the optimal $\text{CRB}^\star$ and $\text{SR}^\star$ can be accordingly calculated.
 \begin{figure*}
  \begin{subequations}
  \begin{align}
    &\mathop {\max }\limits_{{{{\mathbf{\tilde W}}}_i},{{\mathbf{R}}_N},\mathbf{y},z} {\text{  }}\frac{\rho }{{{{\left| {\mathbf{J}} \right|}_{UB}}}}\left| {\mathbf{J}} \right| + \frac{ {1 - \rho } }{{{2^{S{R_{UB}}}}}}z \hfill \\
    &{\text{s}}{\text{.t}}{\text{.  }}{\text{   }}{\text{    }}2c{y_i}\sqrt {{\Sigma _i} + {\text{tr}}\left( {{{{\mathbf{\tilde H}}}_i}{{\mathbf{R}}_N}} \right) + 1}  - y_i^2\left( {{\Sigma _i} - {\text{tr}}\left( {{{{\mathbf{\tilde H}}}_i}{{{\mathbf{\tilde W}}}_i}} \right) + {\text{tr}}\left( {{{{\mathbf{\tilde H}}}_i}{{\mathbf{R}}_N}} \right) + 1} \right) \geq cz, \forall \; i \hfill \\
    &\;\;\;\;\;\;c{\left| {{\alpha _k}} \right|^2}{{\mathbf{a}}^H}\left( {{\vartheta _{k,n}}} \right)\sum\limits_{i = 1}^I {{{{\mathbf{\tilde W}}}_i}} {\mathbf{a}}\left( {{\vartheta _{k,n}}} \right) \leq \left( {1 - c} \right)\left( {{{\left| {{\alpha _k}} \right|}^2}{{\mathbf{a}}^H}\left( {{\vartheta _{k,n}}} \right){{\mathbf{R}}_N}{\mathbf{a}}\left( {{\vartheta _{k,n}}} \right) + 1} \right), \forall\; {\vartheta _{k,n}} \in {\text{card}}\left( {{\Omega _k}} \right),\forall \; k \hfill\\
    &\;\;\;\;\;\;\left( {32b} \right),\left( {32c} \right), \left( {32d} \right) \;\text{and} \; \left( {32e} \right).
  \end{align}
  \end{subequations}
  \rule[-10pt]{18.5cm}{0.05em}
\end{figure*}
 \\\indent To further generalize the problem above and simplify the objective function, we equivalently consider the determinant minimization problem of ${{\mathbf{P}}^H}{\mathbf{J}^{-1}}{\mathbf{P}}$ by introducing the matrix ${\mathbf{P}}$, where $\mathbf{P}$ associates with activated Eves with the dimention of $\mathbf{P}$ is $3K \times 3$. For example, when the CRB minimization is only associated with the first Eve, the first, the $\left( K+1 \right)$-th, and the $\left( 2K+1 \right)$-th rows are the first, second and third rows of the identity matrix ${\mathbf{I}_{3 \times 3}}$, respectively \cite{li2007range}. Then, by noting that the inequality ${\bm{\Upsilon} ^{ - 1}} \geq {{\mathbf{P}}^H}{{\mathbf{J}}^{ - 1}}{\mathbf{P}}$ is equivalent to ${\bm{\Upsilon}}\geq \bm{\Upsilon} {{\mathbf{P}}^H}{{\mathbf{J}}^{ - 1}}{\mathbf{P}}\bm{\Upsilon} $, and based on the Schur-complement condition, problem (32) can be recast as
 \begin{equation}
  \begin{aligned}
    &\mathop {\max }\limits_{{{{\mathbf{\tilde W}}}_i},{{\mathbf{R}}_N},z,\bm{\Upsilon} } \frac{\rho }{{{{\left| {\mathbf{J}} \right|}_{UB}}}}\left| {\bm{\Upsilon}} \right| + \frac{ {1 - \rho } }{{{2^{S{R_{UB}}}}}}z  \hfill \\
    &{\text{s}}{\text{.t}}{\text{.  }}{\text{   }}{\text{    }}\left[ {\begin{array}{*{20}{c}}
      \bm{\Upsilon} &{\bm{\Upsilon} {{\mathbf{P}}^H}} \\
    {{\mathbf{P}}\bm{\Upsilon} }&{\mathbf{J}}
    \end{array}} \right] \succeq {\mathbf{0}} \hfill \\
    &\;\;\;\;\;\;\left( {36b} \right), \left( {36c} \right) \;\text{and} \; \left( {36d} \right). \hfill \\
  \end{aligned}
 \end{equation}
 Similarly, the determinant maximization problem above is convex and readily solvable. For clarity, the above procedure has been summarized in Algorithm 1.

\section{Numerical Results}
In this section, we provide the numerical results to evaluate the effectiveness of the proposed sensing-aided secure ISAC system design. We assume that both the ISAC BS and the radar receiver are equipped with uniform linear arrays (ULAs) with the same number of elements with half-wavelength spacing between adjacent antennas. In the following simulations, the number of transmit antennas and recieve antennas are set as $N_t=N_r=10$ serving $I=3$ CUs, the frame length is set as $L=64$, the noise variance of the communication system is $\sigma _C^2 = 0\;\text{dBm}$.
\\\indent Resultant beampatterns of the proposed sensing-aided ISAC security technique are shown in Fig. 4 and Fig. 5, which demonstrate the single-Eve (located at ${\vartheta _{1,0}}=-25^\circ$) scenario and multi-Eve scenario (located at ${\vartheta _{1,0}}=-25^\circ, {\vartheta _{2,0}}=15^\circ$), respectively. Note that the Rician factor is set as $v_i=0.1$ for generating a Rician channel with weak LoS component, aiming to alleviate the impact on the radar beamppatern caused by the channel correlation, and $\alpha$ is set as $\alpha = 0.05$. To verify the efficiency of the proposed approach, the receive SNR of the echo signal is set as $\text{SNR=-22\;dB}$, which is defined as ${\text{SNR}} = \frac{{{{\left| \beta  \right|}^2}L{P_0}}}{{\sigma _R^2}}$. The ISAC BS first transmits an omnidirectional beampattern for Eve estimation, with the aid of the CAML technique, which is denoted by green dashed lines in Fig. 4 and Fig. 5. It is referred to as the first iteration and the CRB can be accordingly calculated. Then, to ensure that Eves stay within the angle range of main lobes, we design a beampattern with a wide main beam with a beamwith determined by the CRB obtained from the last iteration, which has been elaborated in Setion VI. By updating the CRB iteratively, main lobes get narrow and point to directions of Eves, as illustrated by the rest of the lines in Fig. 4 and Fig. 5. In the simulations, we repeat the weighted optimization problem until the CRB and the secrecy rate both convergence to a local optimum. The beampatterns also indicate that the main beam gain grows with the main lobe width getting narrow. Besides, Fig. 5 shows that the power towards Eves of interest gets lower comparing with single-Eve scenario, while it still outperforms the omnidirectional beampattern design.
\begin{figure}[htbp]
  \centering
  \includegraphics[width=1\columnwidth]{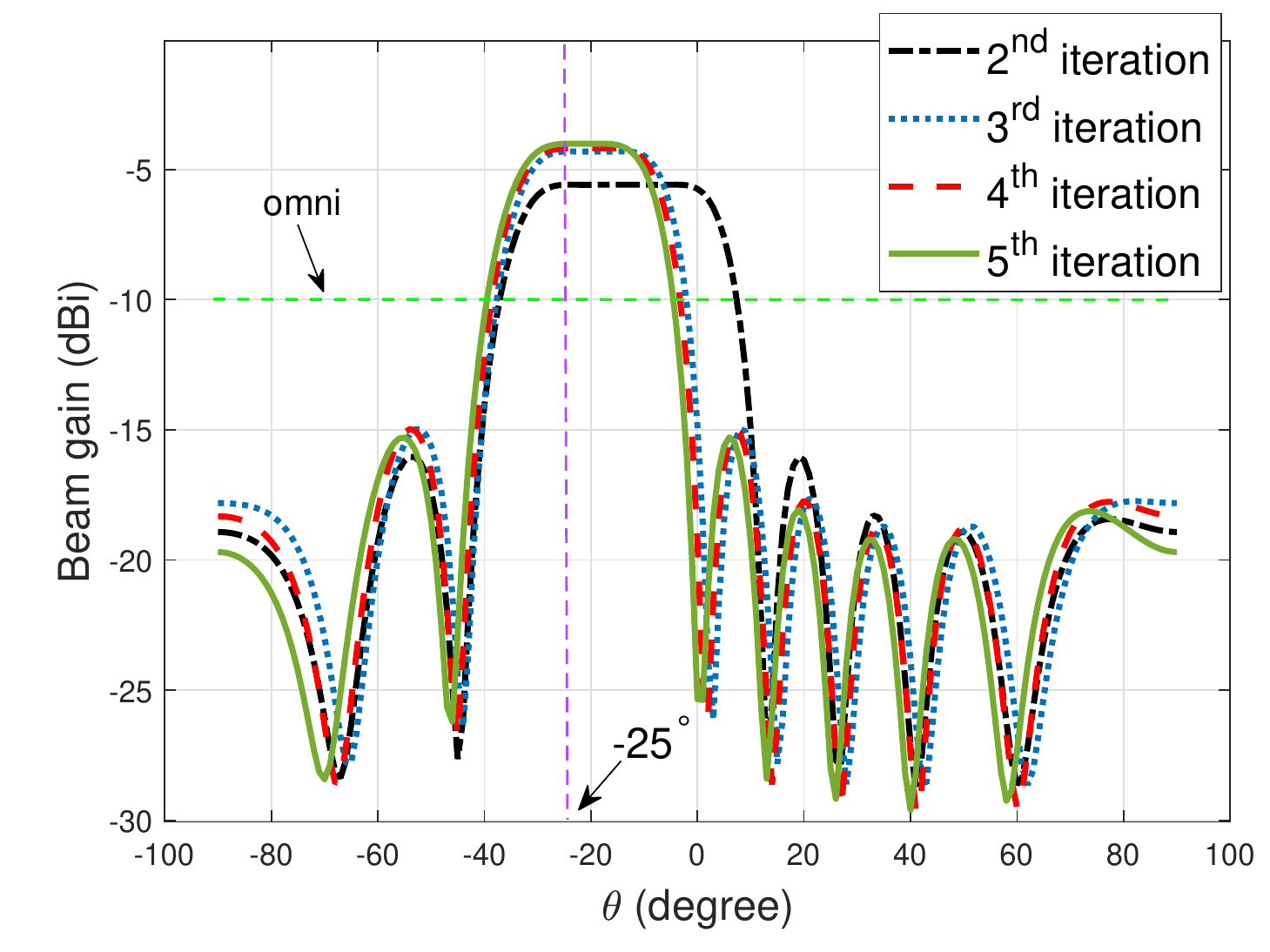}
  \captionsetup{font={footnotesize}}
  \caption{Beampatterns for the scenario of single Eve angle estimation, where the main beam width narrows over each iteration, ${\vartheta _{1,0}}=-25^\circ, I=3, K=1, P_0=\text{35dBm}$, SNR=-22dB.}
  \label{fig.4}
\end{figure}
\begin{figure}[htbp]
  \centering
  \includegraphics[width=1\columnwidth]{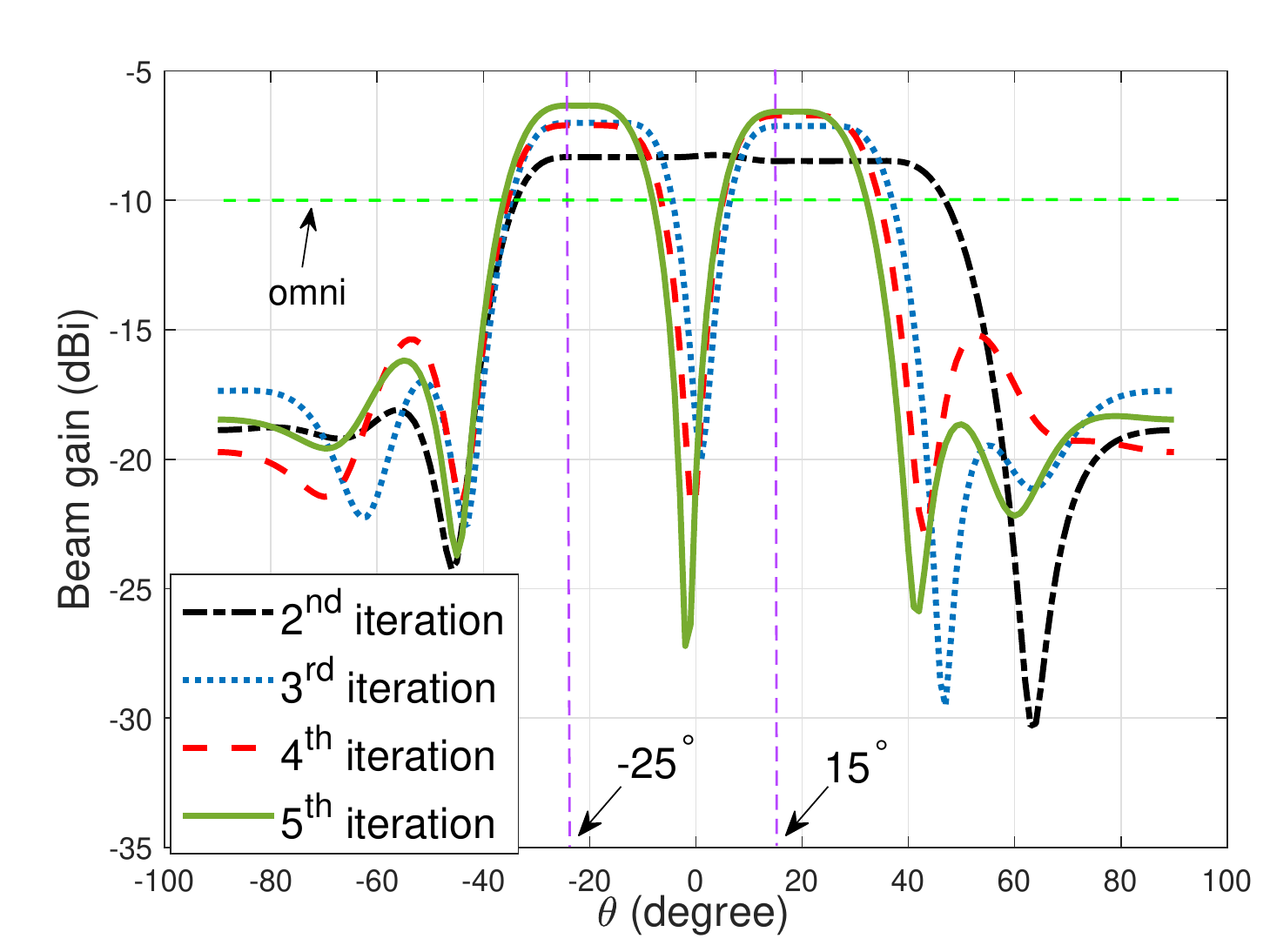}
  \captionsetup{font={footnotesize}}
  \caption{Beampatterns for the scenario of two Eves to be estimated, illustrating the circumstance when the main lobes overlap at the first iteration, ${\vartheta _{1,0}}=-25^\circ, {\vartheta _{2,0}}=15^\circ, I=3, K=2, P_0=35\text{dBm}$, SNR=-22dB.}
  \label{fig.5}
\end{figure}
\begin{figure*}[htbp]
  \centering
  \subfigure[]{
  \includegraphics[width=0.66\columnwidth]{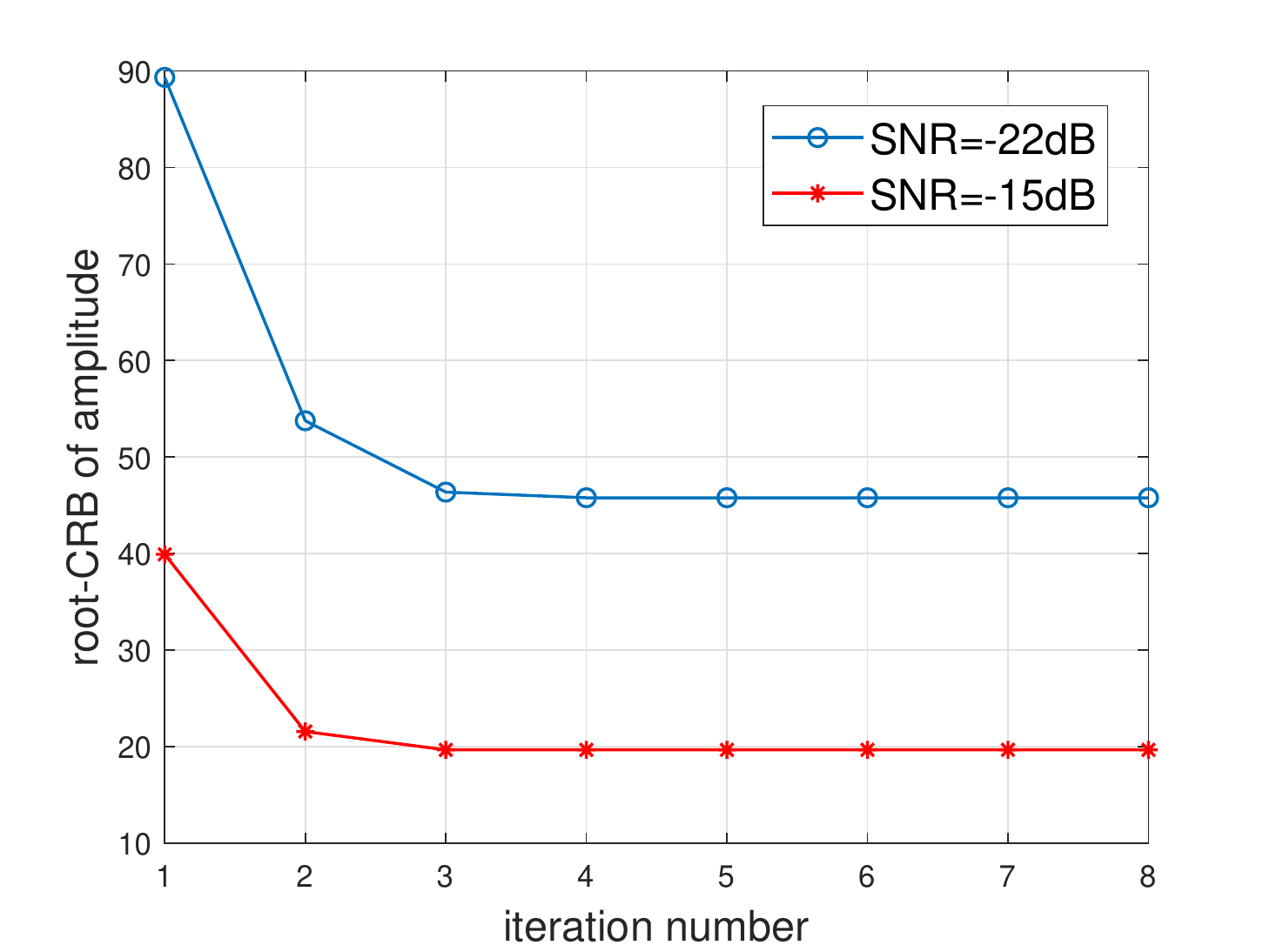}}
  \subfigure[]{
  \includegraphics[width=0.66\columnwidth]{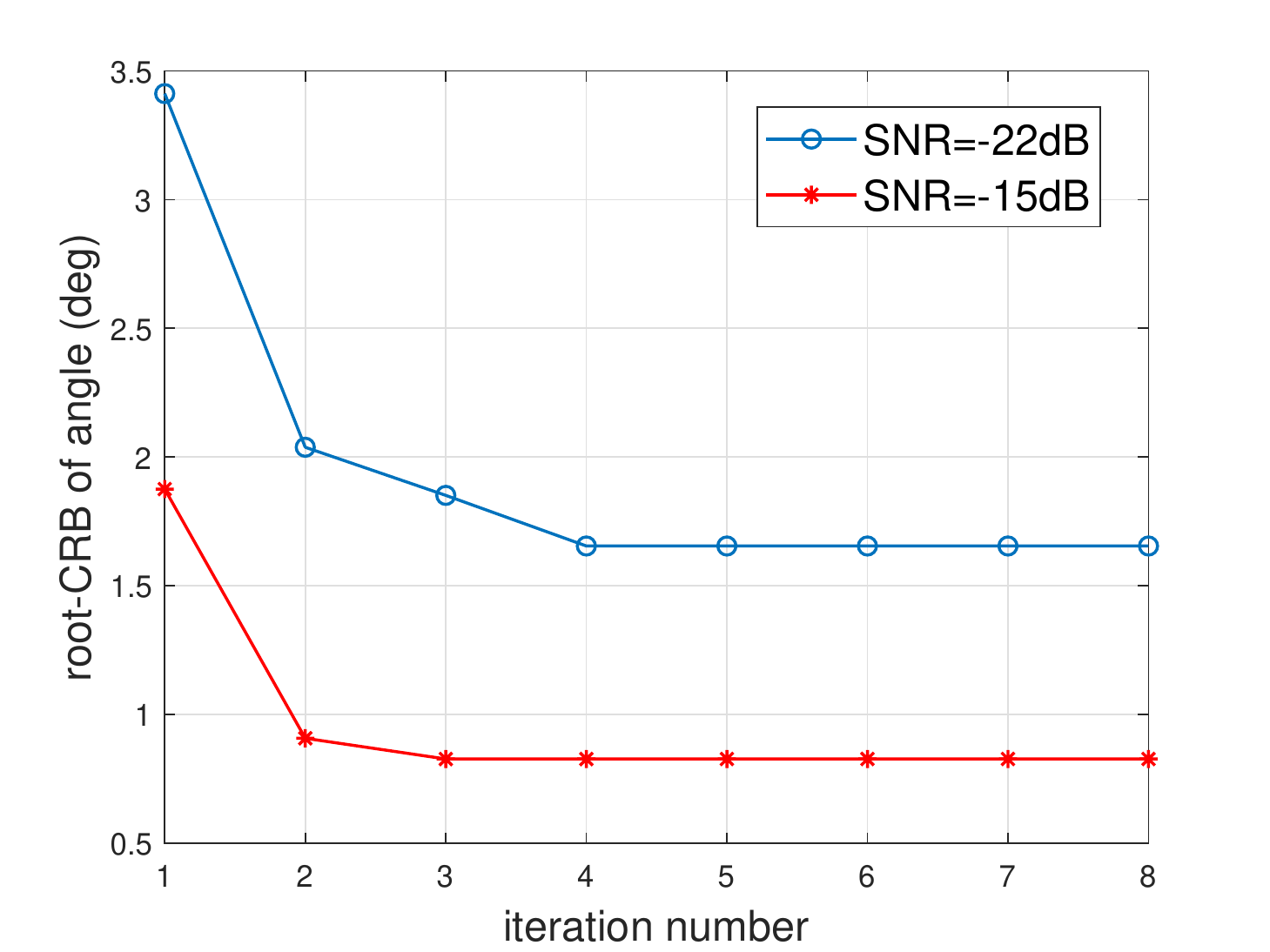}}
  \subfigure[]{
  \includegraphics[width=0.66\columnwidth]{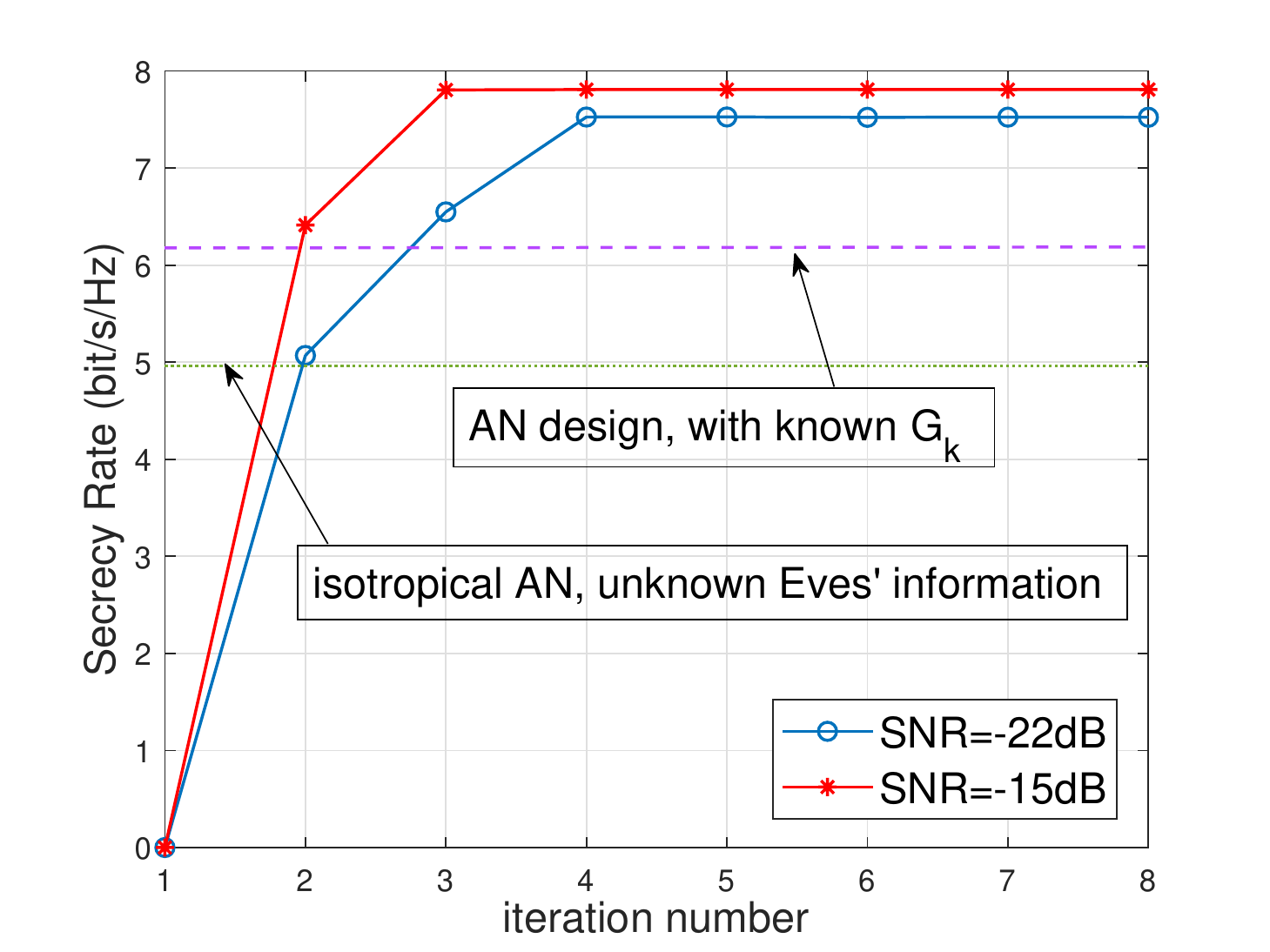}}
  \captionsetup{font={footnotesize}}
  \caption{Convergence with iterations when SNR=-15dB and SNR= 22dB. $I=3, K=1, P_0=35\text{dBm}$. (a) Convergence of root-CRB of amplitude estimation; (b) Convergence of root-CRB of angle estimation; (c) Convergence of the scerecy rate.}
  \label{fig.6}
\end{figure*}
\\\indent Fig. 6 illustrates the convergence of the CRB and the scerecy rate of the proposed algorithm. The benachmark in Fig. 6 (c) is generated following the AN design techniques in Section III, where the covariance of AWGN received by Eves is set as $\sigma_0^2=0\;\text{dBm}$. It is noted that the performance of metrics converges after five iterations when SNR=-22 dB, while the convergence requires less iterations at higher SNR. Additionally, the secrecy rate obtained by the proposed algorithm converges to 7.5 bit/s/Hz and 7.8 bit/s/Hz when SNR=-22 dB and SNR=-15 dB, which outperforms the isotropical AN methods.
\begin{figure}[htbp]
  \centering
  \includegraphics[width=1\columnwidth]{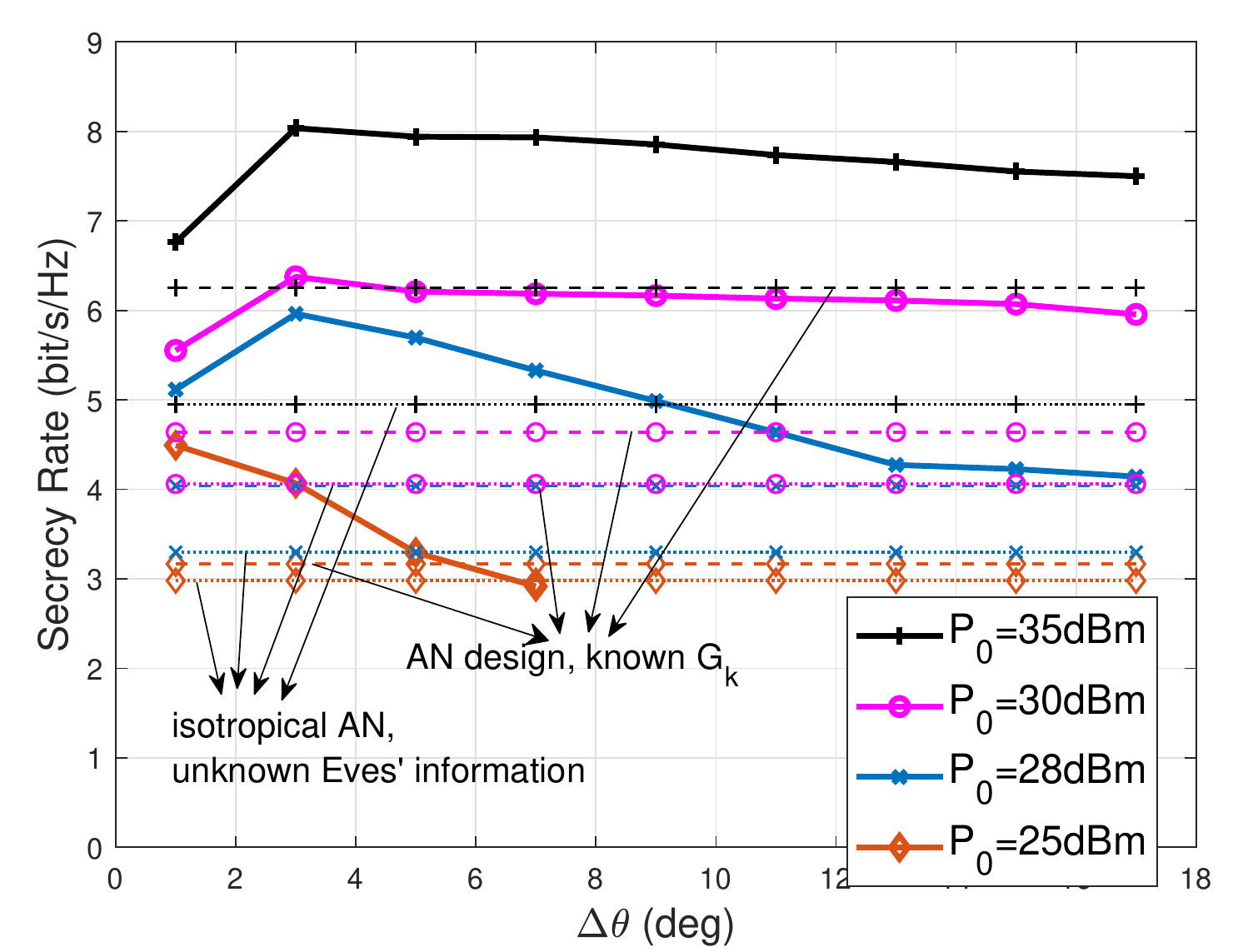}
  \captionsetup{font={footnotesize}}
  \caption{The secrecy rate analysis versus the Eve's location uncertainty with various power budget, where the AN design techniques with no information of Eves' channels and with known $\mathbf{G}_k$ are denoted by dotted lines and dashed lines, respectively. ${\vartheta _{1,0}}=-25^\circ, I=3, K=1, \text{SNR=-15dB}$.}
  \label{fig.7}
\end{figure}
\\\indent In Fig. 7, we investigate the secrecy rate versus the main beam width with different power budget $P_0$, and the benchmarks are given in dashed lines and dotted lines which are obtained by the AN design techniques with knowledge of $\mathbf{G}_k$ and with no information of Eves' channels as given in Sec III, respectively. Generally, the secrecy rate gets higher with the increase of the power budget and it is obvious that the proposed algorithm outperforms benchmark methods. It is worthwhile to stress that the proposed weighted optimization (32) is implemented with no information of Eves. Note that the secrecy rate increases first and then decreases with the expansion of the Eve's location uncertainty. The initial increase is because the gain of the beam towards the target/Eve of interest decreases with the growth of main beam width, resulting in the deterioration of the eavesdropping $\text{SNR}^\text{E}_k$. With respect to the expression in (16), the secrecy rate improves when $\text{SNR}^\text{E}_k$ reduces. However, the power budget constraint becomes tight when the main beam keeps being expanded. This indicates that more power is allocated to the Eve estimation, thus, the secrecy rate decreases. Additionally, when the main beam is wider, the transmission needs to secure the data over a wider range of angles, which is reflected in a SR expression with high channel uncertainty. Particularly, when the power budget is low, for example $P_0=25\;\text{dBm}$, we note that the secrecy rate monotonically decreases with the growth of $\Delta\theta$, while the weighted optimization problem is infeasible due to the power budget limit when the $\Delta\theta$ is larger than 5 degree.
\\\indent Moreover, it is illustrated in Fig. 8 that the secrecy rate decreases with the growth of CU number, given different power budget $P_0$. Note that higher power budget achieves better security performance. Particularly, the secrecy rate cannot be ensured if the ISAC system serves more than 5 CUs when $P_0=25\;\text{dBm}$. In Fig. 9, we consider the performance tradeoff between the target/Eve estimation and communciation data security with different power budget by varying the weighting factor $\rho$. We note that higer $P_0$ results in a better performance of the estimation metric, i.e., root-CRB of the amplitude and the angle. Additionally, with the increase of secrecy rate, the CRB grows as well, which demonstrates the deterioration of the Eve's angle esimation accuracy.
\begin{figure}[htbp]
  \centering
  \includegraphics[width=1\columnwidth]{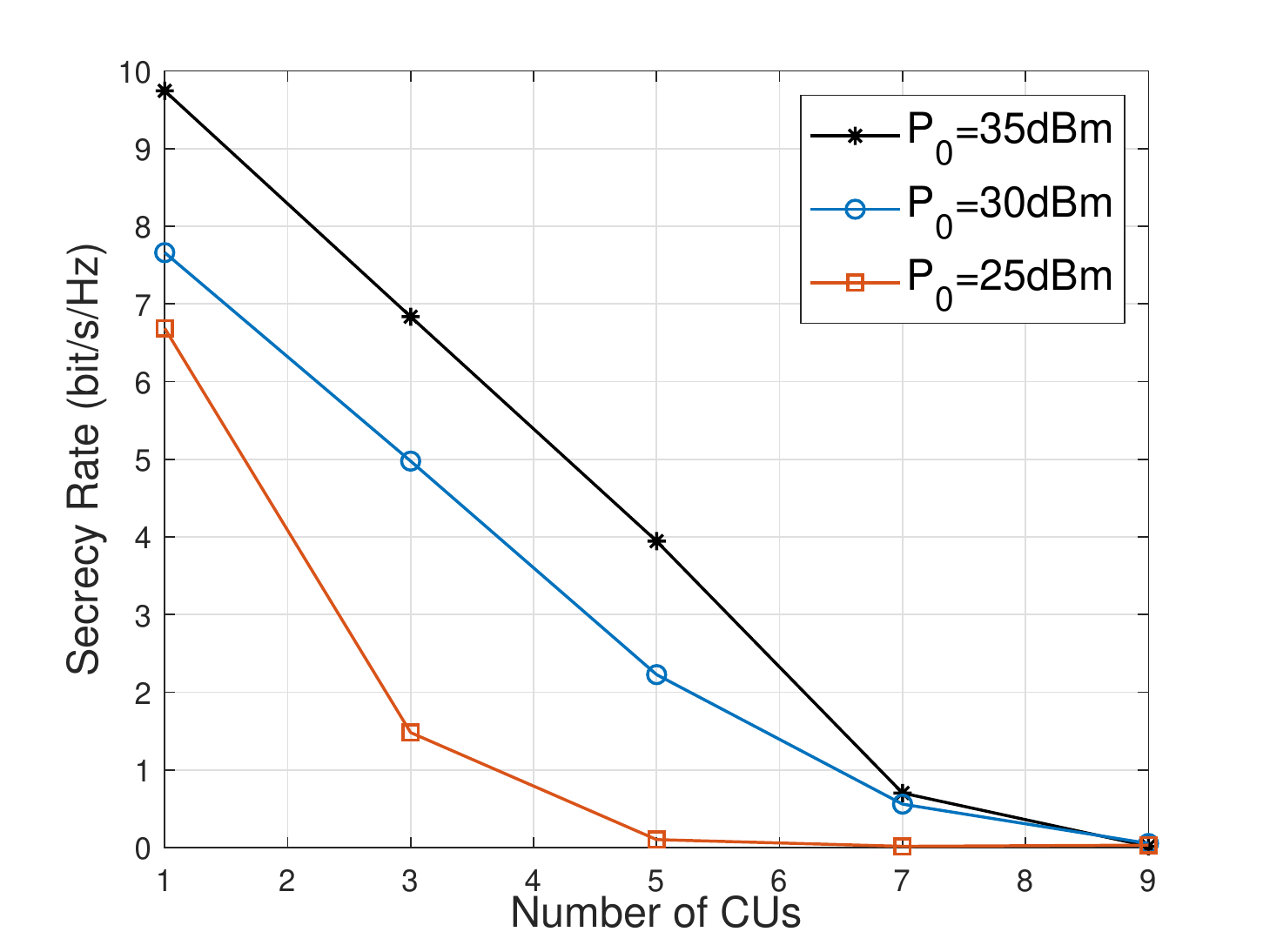}
  \captionsetup{font={footnotesize}}
  \caption{The secrecy rate analysis versus the number of CUs, with various power budget. $K=1, \text{SNR=-15dB}$.}
  \label{fig.8}
\end{figure}
\\\indent Furthermore, we consider a scenario including one CU and one Eve for exploiting impacts on security and sensing metrics resulted from the angle difference between the CU and the Eve. In this case, the Rician channel model with strong LoS component is deployed, i.e., $v_i = 7$ in (2), and the CU is assumed to locate at $-20^\circ$. Resultant beampatterns are shown in Fig. 10 when the Eve is at $-20^\circ$ as well. It is demonstrated that the main beam width converges after four iterations and the generated angle root-CRB at the second iteration is lower than the case of weak Rician channel, which is validated in Fig. 11. Fig.11 illustrates the analysis of the secrecy rate and the root-CRB of angle with various angle difference. Generally speaking, with the expansion of the uncertain angular interval $\Delta \theta$, both of the metrics are deteriorated. The secrecy rate decreases when the Eve and the CU directions get closer, while the performance of the CRB improves since the tradeoff is revealed in Fig. 9.

\begin{figure}[htbp]
  \centering
  \subfigure[]{
  \includegraphics[width=0.48\columnwidth]{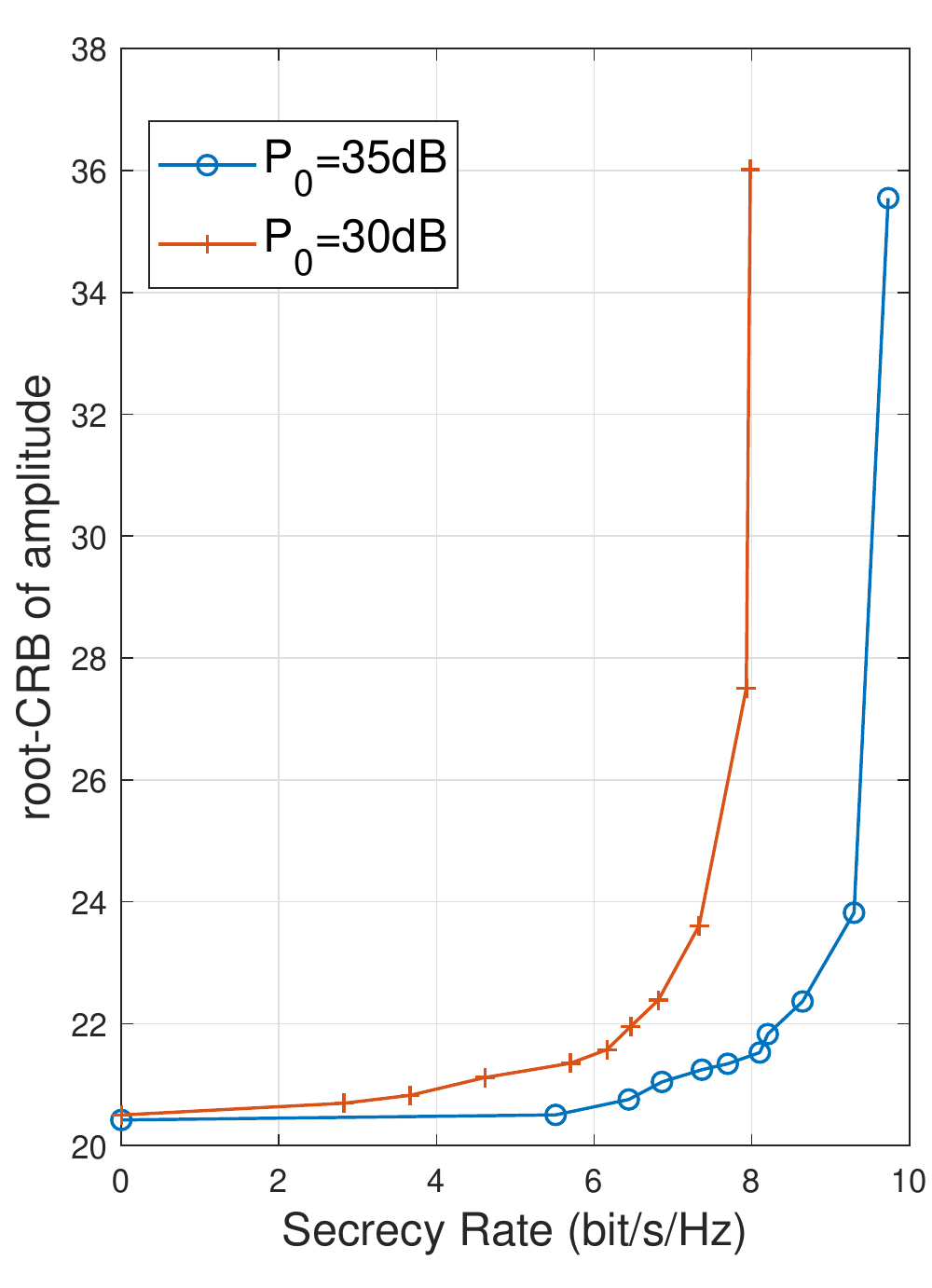}}
  \subfigure[]{
  \includegraphics[width=0.48\columnwidth]{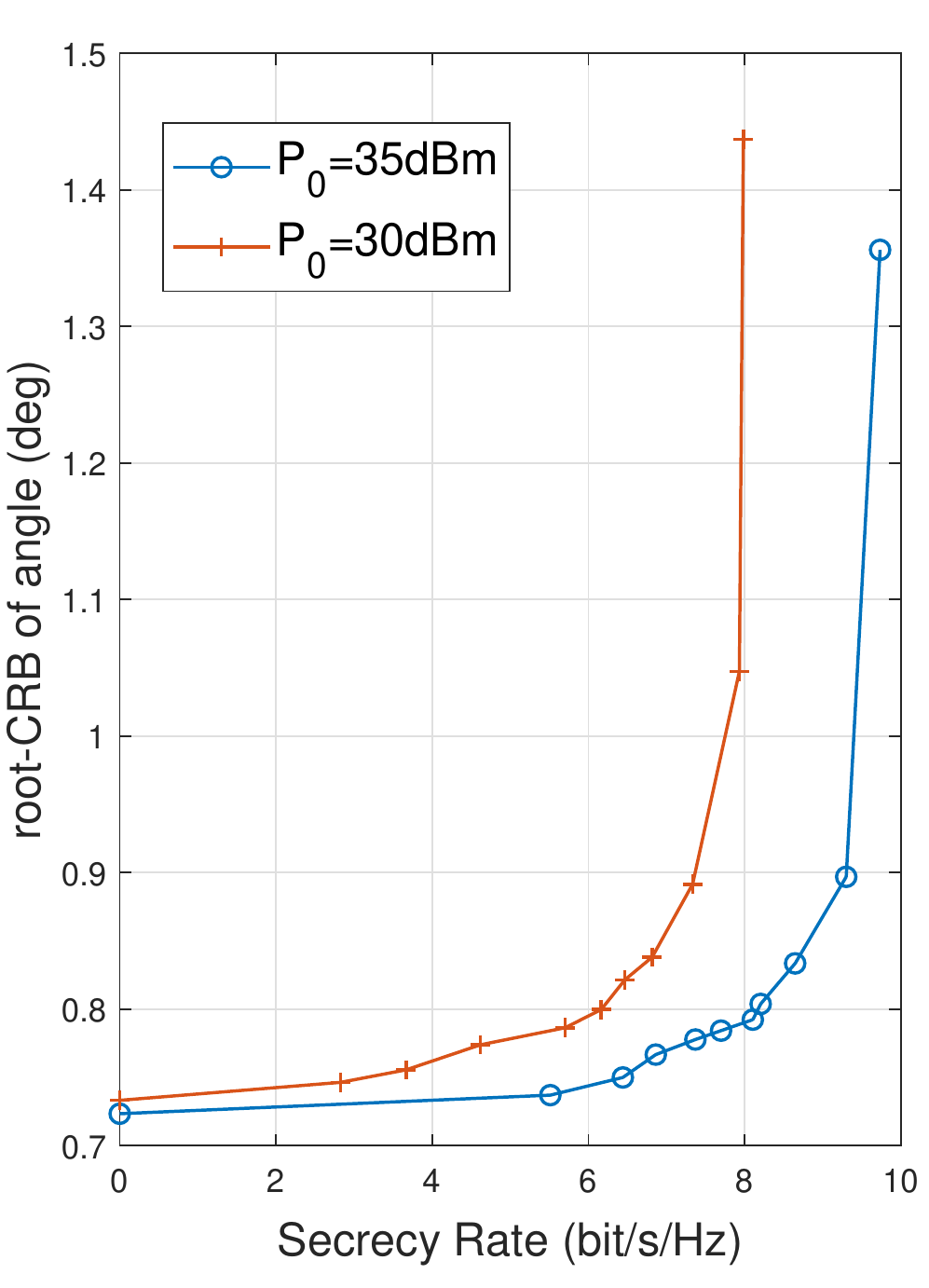}}
  \captionsetup{font={footnotesize}}
  \caption{Tradeoff between the CRB and the secrecy rate with differen power budget. ${\vartheta _{1,0}}=-25^\circ, I=3, K=1, \text{SNR=-15dB}$.}
  \label{fig.9}
\end{figure}

\begin{figure}[htbp]
  \centering
  \includegraphics[width=1\columnwidth]{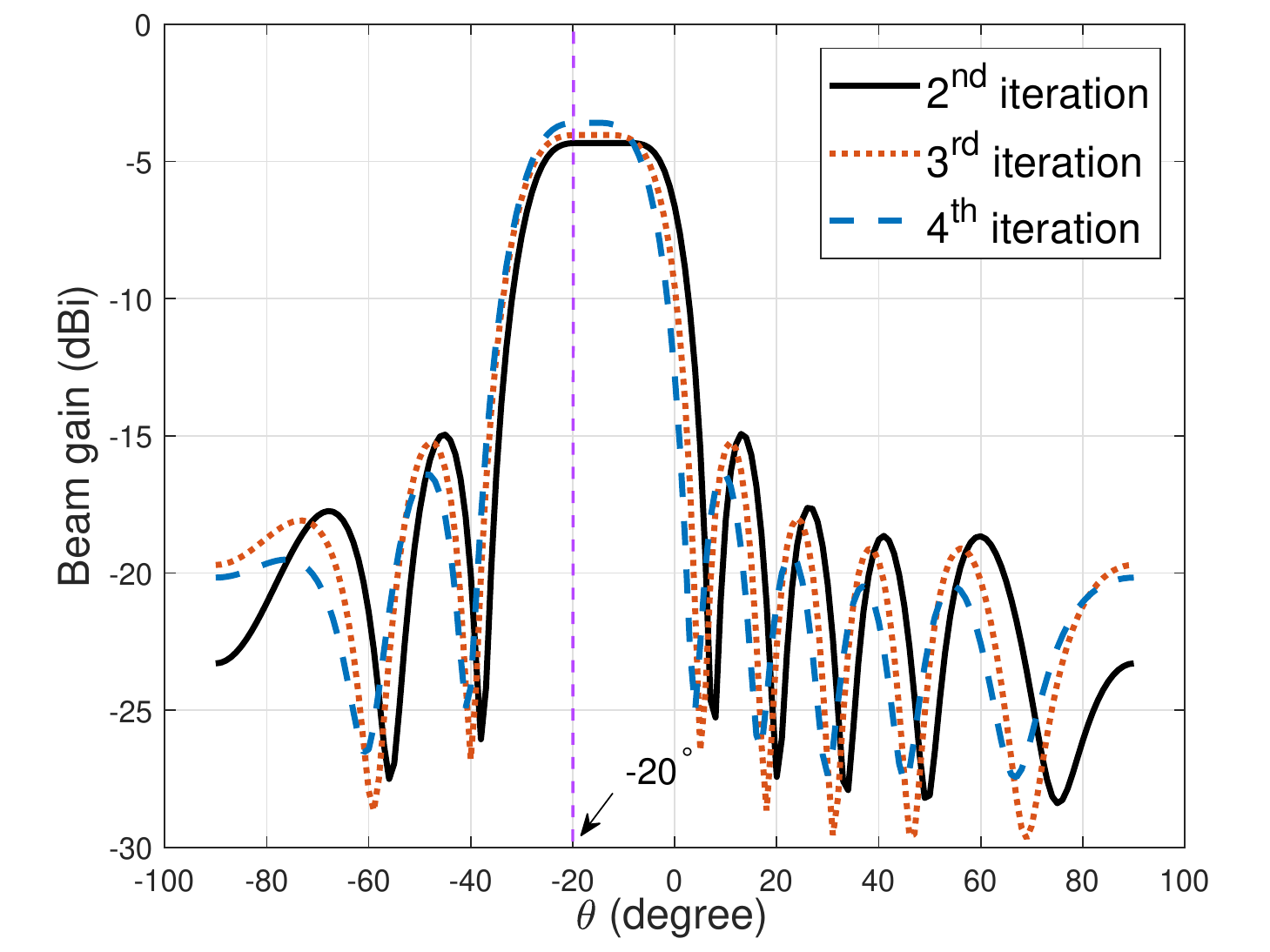}
  \captionsetup{font={footnotesize}}
  \caption{Beampatterns for scenario when the CU and the Eve both locate at $-20^\circ$, narrowing with each iteration until convergence. $I=1, K=1, \text{SNR=-22dB}, P_0=35\text{dBm}$.}
  \label{fig.10}
\end{figure}

\begin{figure}[htbp]
  \centering
  \subfigure[]{
  \includegraphics[width=0.48\columnwidth]{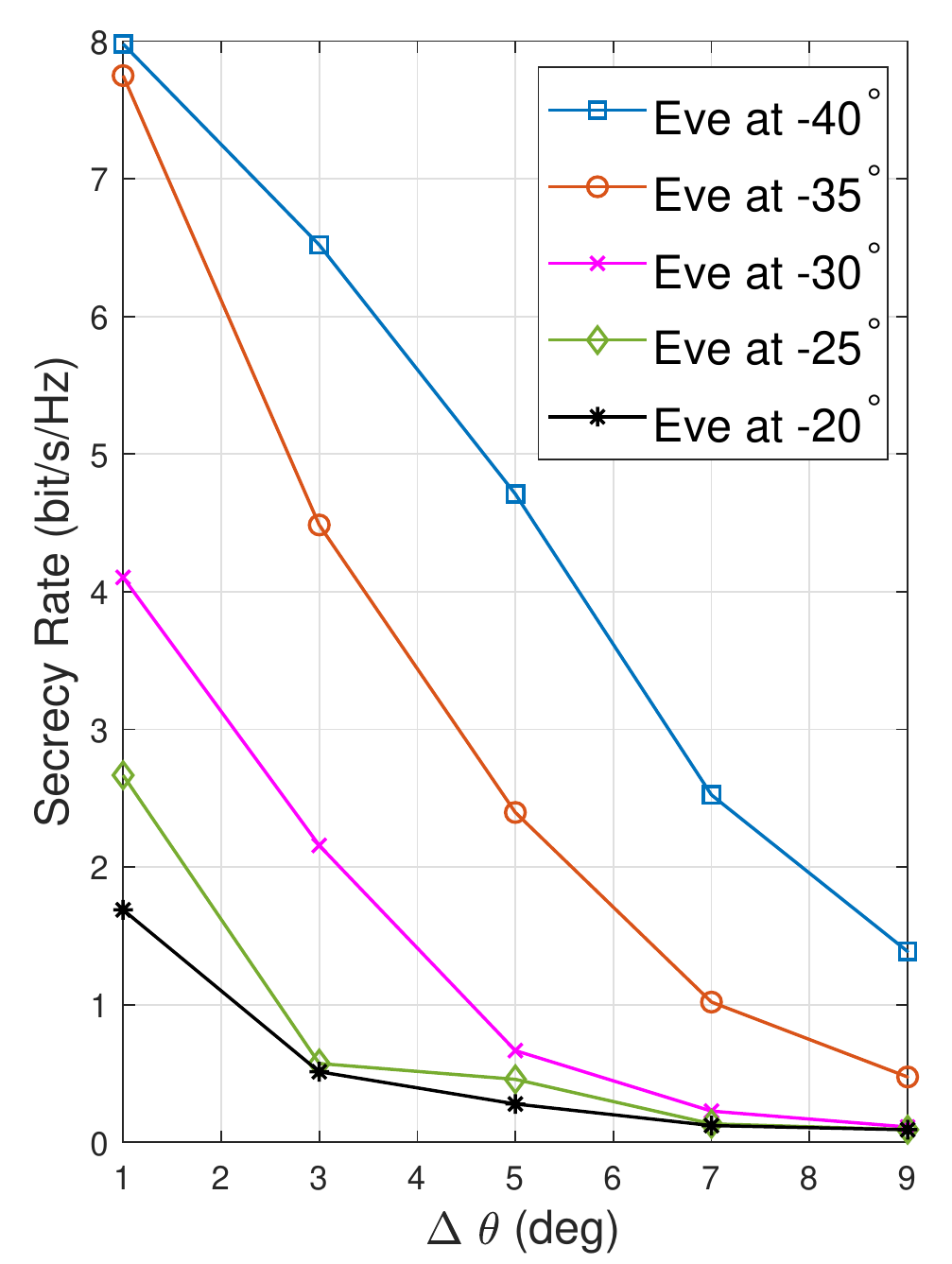}}
  \subfigure[]{
  \includegraphics[width=0.48\columnwidth]{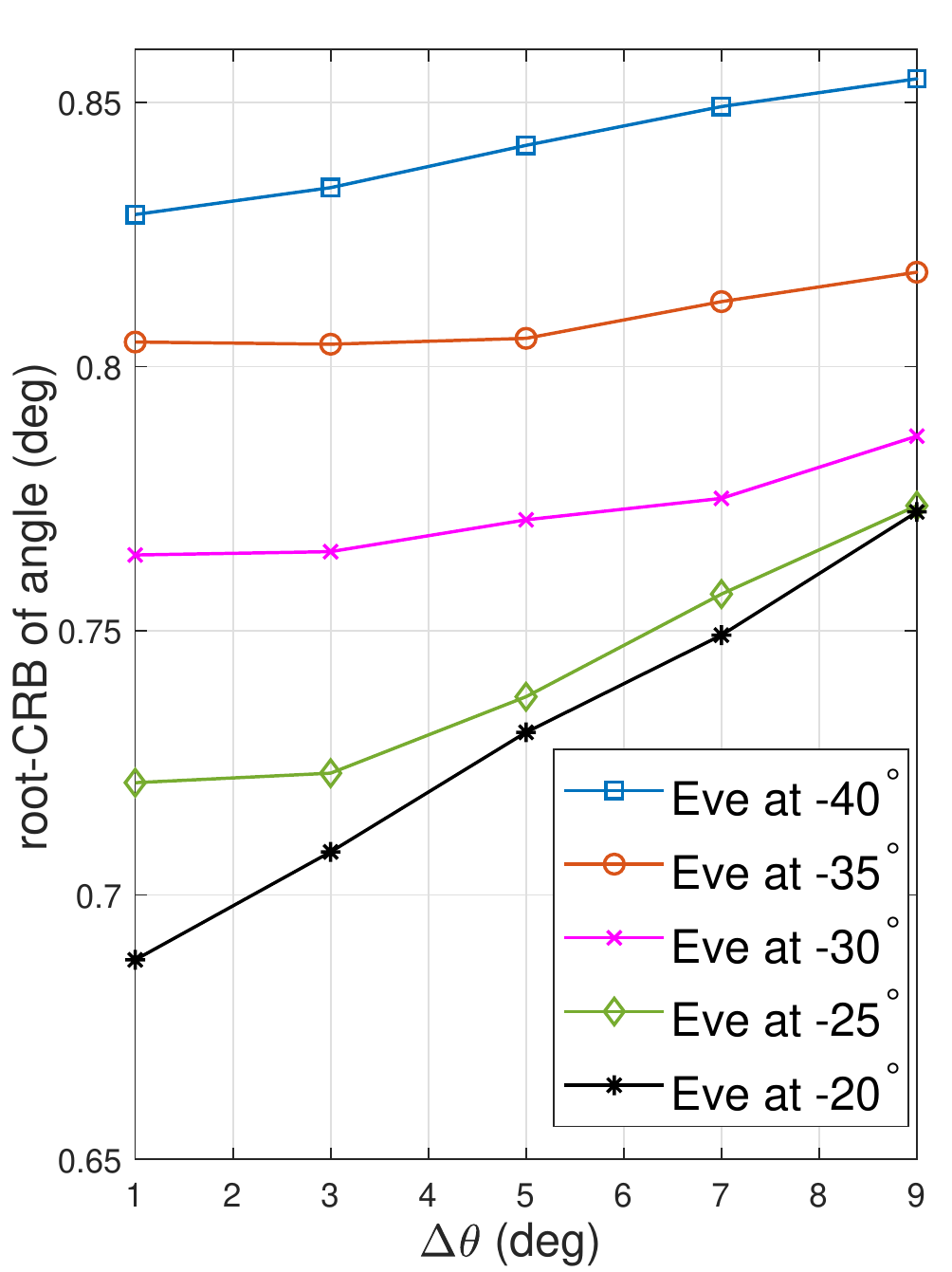}}
  \captionsetup{font={footnotesize}}
  \caption{Secrecy rate and root-CRB of angle performances vesus uncertain angular interval of the target/Eve, with various angle difference between the Eve and the CU, where the CU locates at $-20^\circ$. $I=1, K=1, \text{SNR=-15dB}, P_0 = 35\text{dBm}$.}
  \label{fig.11}
\end{figure}

\section{Conclusion}

In this paper, we have considered the sensing-aided secure ISAC systems, where the dual-functional BS emitted waveforms to estimate the amplitudes and the directions of potential eavesdroppers and send confidential communication data to CUs simultaneously. The proposed design has promoted the cooperation between sensing and communciation rather than conventionally individual functionalities. The weighted optimization problem has been designed to optimize the normalized CRB and secrecy rate while constraining the system power budget. Our numerical results have demonstrated that the secrecy rate was enhanced with the decreasing CRB in both single and multi-Eve scenarios.


\ifCLASSOPTIONcaptionsoff
  \newpage
\fi



\bibliographystyle{IEEEtran}
\bibliography{IEEEabrv,CEP_REF}

\end{document}